\documentclass[12pt]{article}
\usepackage[english]{babel}
\usepackage[dvips]{epsfig}
\usepackage{cite}
\usepackage{color}
\usepackage{colordvi}

\usepackage{amsmath,amssymb}
\usepackage{epsfig}

\newcommand{\ep}{\varepsilon}
\newcommand{\Li}[2]{{\mbox{Li}}_{#1}\left(#2\right)}

\newcommand{\Snp}[2]{{\mbox{S}}_{#1\!}\left(#2\right)}

\begin{document}
\thispagestyle{empty}
\centerline{DESY 13--236 \hfill}
\centerline{December 2013\hfill}
\vspace*{2.0cm}
\begin{center}
{\large \bf
{\sf HYPERDIRE} \\
HYPERgeometric functions DIfferential REduction: \\
MATHEMATICA based packages for
differential reduction of generalized hypergeometric functions: \\
$F_D$ and $F_S$ Horn-type hypergeometric functions of three variables
}
\end{center}
\vspace*{0.8cm}

\begin{center}
{\sc Vladimir~V.~Bytev$^{a}$}, \quad
{\sc Mikhail~Yu.~Kalmykov$^{b}$}, \quad
{\sc Sven-Olaf~Moch$^{b,c}$\footnote{{\bf e-mail}: sven-olaf.moch@desy.de} } \\
 \vspace*{0.8cm}

{\normalsize $^{a}$ Joint Institute for Nuclear Research,} \\
{\normalsize $141980$ Dubna (Moscow Region), Russia} \\[2ex]
{\normalsize $^{b}$ Deutsches Elektronen-Synchrotron DESY}\\ 
{\normalsize Platanenallee 6, 15738 Zeuthen, Germany} \\[2ex]
{\normalsize $^{c}$ II. Institut f\"ur Theoretische Physik, Universit\"at Hamburg,}\\
{\normalsize Luruper Chaussee 149, 22761 Hamburg, Germany} \\
\end{center}
\vspace*{0.8cm}

\begin{abstract}
{\bf HYPERDIRE} is a project devoted to the creation of a set of {\tt Mathematica} 
based programs for the differential reduction of hypergeometric functions.
The current version includes two parts:
the first one, {\bf FdFunction}, 
for manipulations with Appell hypergeometric functions $F_D$ of $r$ variables; 
and the second one, {\bf FsFunction}, 
for manipulations with Lauricella-Saran hypergeometric functions $F_S$ of three variables. 
Both functions are related with one-loop Feynman diagrams.

\medskip


\end{abstract}

\newpage

{\bf\large PROGRAM SUMMARY}
\vspace{4mm}
\begin{sloppypar}
\noindent   {\em Title of program\/}: {\sf HYPERDIRE} \\[2mm]
   {\em Version\/}: 1.0.0
   {\em Release\/}: 1.0.0
   {\em Catalogue number\/}: \\[2mm]
   {\em Program obtained from\/ {\tt https://sites.google.com/site/loopcalculations/home}}:
   {\tt } \\[2mm]
   {\em E-mail: bvv@jinr.ru} \\[2mm]
   {\em Licensing terms\/}: GNU General Public License  \\[2mm]
   {\em Computers\/}: all computers running {\tt Mathematica} \\[2mm]
   {\em Operating systems\/}:  operating systems running {\tt Mathematica}\\[2mm]
   {\em Programming language\/}: {\tt Mathematica} \\[2mm]
   {\em Nature of the problem\/}:
                  Reduction of hypergeometric functions $F_D$ and $F_S$ to set of basis functions.
                \\[2mm]
   {\em Method of solution\/}: Differential reduction \\[2mm]
   {\em Restriction on the complexity of the problem}: none \\[2mm]
   {\em Typical running time}:  Depending on the complexity of problem.
\end{sloppypar}
%
\newpage

{\bf\large LONG WRITE-UP}
\vspace{4mm}

\section{Introduction}
The study of solutions of linear partial 
differential equations (PDEs) of a few variables in terms of multiple series,
i.e., a multivariable generalization of Gauss hypergeometric function~\cite{Gauss}, 
was started a long time ago \cite{Lauricella}. 
Following the Horn definition~\footnote{The modern approach 
to hypergeometric functions has been presented in  \cite{Gelfand}.}, 
a multiple series is called Horn-type hypergeometric function \cite{multiple}, 
if around some point $\vec{z}=\vec{z_0}$, there are series representations
$$
H(\vec{z}) = \sum_{\vec{m}} C(\vec{m}) \left(\vec{z} \!-\! \vec{z_0} \right)^{~\vec{m}},
$$
where ${\vec{m}}$ is a set of integers 
and the ratio of two coefficients can be represented as a ratio of
two polynomials:
\begin{equation}
\frac{C(\vec{m}+\vec{e}_j)}{C(\vec{m})}  =  \frac{P_j(\vec{m})}{Q_j(\vec{m})} 
\;,
\label{horn}
\end{equation}
where
$                                                                                            
\vec{e}_j = (0,\cdots,0,1,0,\cdots,0),                                                       
$
is the $j^{\rm th}$ unit vector. 
The coefficients $C(\vec{m})$ of such a series are expressible as 
product/ratio of Gamma-functions (up to some factors irrelevant for our consideration)
\cite{Ore:Sato}:
\begin{eqnarray}                                                                             
C(\vec{m})                                                                                   
= 
\frac{
\prod\limits_{j=1}^p
\Gamma\left(\sum_{a=1}^r \mu_{ja}m_a+\gamma_j \right)}
{
\prod\limits_{k=1}^q
\Gamma\left( \sum_{b=1}^r \nu_{kb}m_b+\sigma_k \right)
} \;,
\label{ore}
\end{eqnarray}
where
$                                                                                            
\mu_{ja}, \nu_{kb}, \sigma_j,\gamma_j \in \mathbb{Z}
$
and $m_a$ are elements of ${\vec{m}}$.

The Horn-type hypergeometric function, Eq.~(\ref{horn}),  
satisfies the following system of differential equations:
\begin{equation}
0 =
D_j (\vec{z})
H(\vec{z})
=
\left[
Q_j\left(
\sum_{k=1}^r z_k\frac{\partial}{\partial z_k}
\right)
\frac{1}{z_j}
-
 P_j\left(
\sum_{k=1}^r z_k\frac{\partial}{\partial z_k}
\right)
\right]
H(\vec{z}) \;,
\label{diff}
\end{equation}
where $j=1, \ldots, r$.
The degree of polynomials $P_i$ and $Q_i$ is $p_i$ and $q_i$, respectively.
The largest of these numbers, $r=\max \{p_i,q_j \}$, is called the order of the hypergeometric series. 

Any Horn-type hypergeometric function is a function of two kind of variables, 
{\it continuous} variables, $z_1,z_2, \cdots, z_r$ and 
{\it discrete} variables: $\{ J_a \}:= \{\gamma_k,\sigma_r \}$, 
where the latter can change by integer numbers 
and are often referred to as the {\it parameters} of the hypergeometric function. 
For any Horn-hypergeometric function,  there are linear differential operators 
changing the value of the discrete variables by one unit: 
\begin{eqnarray}
R_{K}(\vec{z})\frac{\partial^K }{\partial \vec{z}} H(\vec{J};\vec{z}) 
= H(\vec{J} \pm e_K; \vec{z}) \;,
\label{direct}
\end{eqnarray}
where $R_{K}(\vec{z})$ are polynomial (rational) functions. 
In Refs.\ \cite{miller,algorithm} it was shown that there is an algorithmic solution 
for the construction of inverse linear differential operators: 
%
%
%
\begin{eqnarray}
B_{L}(\vec{z})\frac{\partial^L }{\partial \vec{z}} 
\left( R_{K}(\vec{z})\frac{\partial^K }{\partial \vec{z}} \right) 
H(\vec{J};\vec{z}) 
\equiv 
B_{L}(\vec{z})\frac{\partial^L }{\partial \vec{z}} H(\vec{J} \pm e_K; \vec{z}) 
= 
H(\vec{J};\vec{z}) 
\;.
\label{inverse}
\end{eqnarray}

Applying the direct or inverse differential operators to the hypergeometric function,
the value of parameters can be changed by an arbitrary integer number: 
\begin{equation}
S(\vec{z}) H(\vec{J}+\vec{m}; \vec{z})
= 
\sum_{j=0}^r S_j(\vec{z}) 
\frac{\partial^j}{\partial \vec{z}} H(\vec{J}; \vec{z}) \;,
\label{reduction}
\end{equation}
where $\vec{m}$ is a set of integers, $S$ and $S_j$ are polynomials 
and $r$ is the holonomic rank (the number of linearly independent solutions) 
of the system of differential equations, Eq.~(\ref{diff}).
Additionally, the construction of inverse differential operators defined 
by Eq.~(\ref{inverse}) allows to 
\begin{itemize}
\item[(i)] 
find a set of exceptional parameters for any hypergeometric function, 
and this set coincides with the condition of reducibility 
of the monodromy group of the corresponding hypergeometric functions
(see discussion in \cite{hyperdire:1} and Section \ref{EXCEP} for $F_D$ and $F_S$ functions);
\item [(ii)] 
convert the system of linear PDEs, Eq.~(\ref{diff}), into Pfaff form for any hypergeometric functions,
including functions with Puiseux monomials as one of the solution, see details in \cite{hyperdire:2b}. 
\end{itemize}
The interest of physicists in hypergeometric functions is related with 
\begin{itemize}
\item[(i)] 
the necessity of an analytical evaluation of multiple series generated by multiple residues of Mellin-Barnes integrals \cite{smirnov-tausk};
\item [(ii)] 
the restricted set of values of parameters of hypergeometric functions 
or multiple series, where the algorithms \cite{nested,nested2,AS,veretin} are applicable; 
\item [(iii)] 
the complicated analytical structure of one-loop massive Feynman diagrams, 
where, nevertheless, a simple hypergeometric representation exists \cite{fjt,dd,kt}. 
\end{itemize}
It was pointed out in \cite{expansion:1} that the differential reduction algorithm, 
defined as a full system of differential operators, Eqs.~(\ref{diff}), (\ref{direct}), (\ref{inverse}), 
can be applied to the construction of analytical coefficients of the so-called $\varepsilon$-expansions 
of hypergeometric functions about any rational values of parameters 
via the direct solution of the linear systems of differential equations.

This is the motivation for creating a package for the manipulation of the parameters of
Horn-type hypergeometric functions of several variables.

In the previous publications the algebraic reduction of ${}_2F_1$ functions has been considered~\cite{hyperdire},
the program {\bf pfq}, 
for the manipulation of hypergeometric functions, $_{p+1}F_{p}~(p \geq 1)$~\cite{hyperdire:1}, 
the program {\bf AppellF1F4}, for the manipulation of 
Appell hypergeometric functions, $F_1,F_2,F_3$ and $F_4$~\cite{hyperdire:2a}, 
the program~{\bf Horn}, for the manipulation of Horn-hypergeometric functions 
of two variables (30 hypergeometric functions in addition 
to four Appell functions) \cite{hyperdire:2b}.

The aim of this paper is to present a further extension of the {\tt Mathematica} \cite{math} based package {\bf HYPERDIRE} 
for the differential reduction of the Horn-type hypergeometric
function with arbitrary values of parameters to a set of basis functions.
The current version consists of two parts:
one, {\bf FdFunction}, for the manipulation of 
Lauricella hypergeometric functions, $F_D$, of $r$ variables,
and the second one, {\bf FsFunction}, for the manipulation with 
Lauricella-Saran hypergeometric functions $F_S$ with three variables.

\section{The structure of hypergeometric functions related with 
  one-loop off-shell Feynman diagrams}

A generic scalar one-loop $N$-point function is defined by the following integral 
in $d$ space-time dimensions
\begin{eqnarray}
{\lefteqn{
I_{N;a_1,\cdots, a_N}^{(d)} = }}
\nonumber \\ &&
\int\, \frac{d^dl}{(2\pi)^d}\, 
\frac{1}{((l-p_1)^2-m_{12}^2)^{a_1}\,\dots\,((l-(p_1+\dots p_{N-1}))^2-m_{N-1,N}^2)^{a_{N-1}}\, (l^2-m_{N,1}^2)^{a_N}} 
\, ,
\qquad
\label{zero}
\end{eqnarray}
where $l$ is the loop momentum to be integrated, 
$p_i$ are the external momenta and $m_{i,j}^2$ the masses of the internal
propagators, $i,j=1,\dots,N$.
Energy-momentum conservation enforces $\sum_i p_i = 0$.

\subsection{Massive case}
In accordance with the algorithm described in \cite{tarasov}, one-loop $N$-point
diagrams with all powers of propagators equal to unity, 
i.e., all $a_i=1$ in Eq.~(\ref{zero}), 
satisfy to the following difference equation 
\begin{eqnarray}
I_N^{(d)} = b_N(d)
+ \sum_{k=1}^N \left( \frac{\partial_k \Delta_N}{2\Delta_N} \right)
\sum_{r=0}^\infty
\left( \frac{d-N+1}{2} \right)_r
\left( \frac{G_{N-1}}{\Delta_N} \right)^r
{\bf k^-} I_{N}^{(d+2r)} \;,
\label{tarasov}
\end{eqnarray}
where 
$
I_N^{(d)}
\equiv 
I_{N;a_1,\cdots, a_N}^{(d)}, 
$
$(a)_k$ is a Pochhammer symbol, $(a)_k = \Gamma(a+k)/\Gamma(a)$,
the operator ${\bf k^-}$ shifts the value of index $a_k \to a_k-1$ 
(in our case, $a_k \equiv 1 \Longrightarrow {\bf k^-} a_k=0$),
$d$ is the dimension of space-time, 
and $b_N(d)$ is an arbitrary periodic function depending on the kinematic invariants. 
Moreover, we are working in Euclidean space-time, 
which is the source of the sign ``+'' instead of ``-'' as it was defined in \cite{tarasov},
$G_N$ is a Gram determinant, $\Delta_N$ is a Cayley determinant for the $N$-point diagram 
and
$
\partial_k \Delta_N = \frac{\partial}{\partial m_k^2}\Delta_N \;.
$
For details we refer to~\cite{tarasov,fjt}. 
Eq.~(\ref{tarasov}) can be solved iteratively~\cite{tarasov,fjt} and the result 
for the one-loop $N$-point diagram in an arbitrary dimension $d$ can be
written as linear combinations of the following hypergeometric functions: 
\begin{eqnarray}
I_{N \geq 2}^{(d)} & \sim & 
\prod\limits_{j=2}^N
\left( \frac{\partial_{k_j} \Delta_j}{2\Delta_j}  \right)
\times 
\sum_{r_1,r_2,r_3,\cdots,r_{N-1}=0}^\infty 
\left( -m^2_{N-1,N} \frac{G_{N-1}}{\Delta_N} \right)^{r_{N-1}} 
\cdots 
\left( -m^2_{12} \frac{G_1}{\Delta_2} \right)^{r_1} 
\nonumber \\ && 
\times 
\frac{\Gamma\left(\frac{d-N+1}{2} \!+\! r_{N-1} \right)}
     {\Gamma\left(\frac{d-N+1}{2} \right)}
\cdots 
\frac{\Gamma\left(\frac{d-4}{2} \!+\! r_4 \cdots \!+\! r_{N-1} \right)}
     {\Gamma\left(\frac{d-4}{2} \!+\! r_5 \cdots \!+\! r_{N-1} \right)}
\frac{\Gamma\left(\frac{d-3}{2} \!+\! r_3 \!+\! r_4 \cdots \!+\! r_{N-1} \right)}
     {\Gamma\left(\frac{d-3}{2} \!+\! r_4 \cdots \!+\! r_{N-1} \right)}
\nonumber \\ && 
\times 
\frac{\Gamma\left(\frac{d-2}{2} \!+\! r_2 \!+\! r_3 \cdots \!+\! r_{N-1} \right)}
     {\Gamma\left(\frac{d-2}{2} \!+\! r_3 \cdots \!+\! r_{N-1} \right)}
\frac{\Gamma\left(\frac{d-1}{2} \!+\! r_1 \!+\! r_2 \cdots \!+\! r_{N-1} \right)}
     {\Gamma\left(\frac{d-1}{2} \!+\! r_2 \cdots \!+\! r_{N-1} \right)}
\frac{\Gamma\left(\frac{d}{2}\right)}
     {\Gamma\left(\frac{d}{2} \!+\! r_1 \!+\! r_2 \cdots \!+\! r_{N-1} \right)}
\nonumber \\ && 
+   \sum_{j=3}^N b_j(d) c_j(d)
\;,
\label{N}
\end{eqnarray}
where $m^2_{i,j}$ are some masses, cf., Eq.~(\ref{zero}). 
In accordance with {\bf Proposition 1} of \cite{KK2012}~\footnote{
For completeness, we recall it here: 
A multiple Mellin-Barnes integrals can be presented as 
a linear combination of Horn-type hypergeometric functions about some point.
Therefore, the holonomic rank of the corresponding system of linear differential 
equations related with the Mellin-Barnes integral is equal to the holonomic rank of any 
hypergeometric function in the corresponding hypergeometric representation.
This statement has been  proven only for one-fold Mellin-Barnes integrals.
}, 
the holonomic rank of the product $b_j(d) c_j(d)$, 
is expected to be equal to the holonomic rank of the hypergeometric function, 
defined by Eq.~(\ref{HYPERA}).
Dropping all irrelevant factors, 
the hypergeometric function related with one-loop $N$-point off-shell massive 
Feynman diagram is 
(it has a simpler form in contrast to the results of \cite{davydychev}):
\begin{eqnarray}
H_{N \geq 2}^{(d)} & = & 
\sum_{r_1,r_2,r_3,\cdots,r_{N-1}=0}^\infty 
\frac{\Gamma\left(\frac{d-1}{2} \!+\! r_1 \!+\! r_2 \cdots \!+\! r_{N-1} \right)}
     {\Gamma\left(\frac{d}{2}   \!+\! r_1 \!+\! r_2 \cdots \!+\! r_{N-1} \right)}
\nonumber \\ && 
\times 
\frac{\Gamma\left(\frac{d-2}{2} \!+\! r_2 \!+\! r_3 \cdots \!+\! r_{N-1} \right)}
     {\Gamma\left(\frac{d-1}{2} \!+\! r_2 \cdots \!+\! r_{N-1} \right)}
\frac{\Gamma\left(\frac{d-3}{2} \!+\! r_3 \!+\! r_4 \cdots \!+\! r_{N-1} \right)}
     {\Gamma\left(\frac{d-2}{2} \!+\! r_3 \cdots \!+\! r_{N-1} \right)}
\nonumber \\ && 
\times 
\frac{\Gamma\left(\frac{d-4}{2} \!+\! r_4 \cdots \!+\! r_{N-1} \right)}
     {\Gamma\left(\frac{d-3}{2} \!+\! r_4 \cdots \!+\! r_{N-1} \right)}
\cdots 
\frac{\Gamma\left(\frac{d-N+1}{2} \!+\! r_{N-1} \right)}
     {\Gamma\left(\frac{d-N+2}{2} \!+\! r_{N-1} \right)}
z_1^{r_1} z_2^{r_2} \cdots z_{N-1}^{r_{N-1}} \;.
\label{HYPERA}
\end{eqnarray}
For the lowest values of $N=2,3,4,5,$ Eq.~(\ref{HYPERA}) has the following form:
\begin{eqnarray}
H_2^{(d)} & = & 
\sum_{r_1} 
\frac{\left( \frac{d-1}{2} \right)_{r_1}}{\left(\frac{d}{2} \right)_{r_1}} z_1^{r_1}
\;, 
\label{H2}
\\ 
H_3^{(d)} & = & \sum_{r_1,r_2} \frac{\left( \frac{d-1}{2} \right)_{r_1+r_2} \left( \frac{d-2}{2} \right)_{r_2} }
                              {\left(\frac{d}{2} \right)_{r_1+r_2} \left( \frac{d-1}{2} \right)_{r_2} } 
z_1^{r_1} z_2^{r_2}
\;, 
\label{H3}
\\
H_4^{(d)} & = & \sum_{r_1,r_2,r_3} \frac{\left( \frac{d-1}{2} \right)_{r_1+r_2+r_3} 
                               \left( \frac{d-2}{2} \right)_{r_2+r_3} 
                               \left( \frac{d-3}{2} \right)_{r_3} 
                              }
                              {\left(\frac{d}{2} \right)_{r_1+r_2+r_3} 
                               \left( \frac{d-1}{2} \right)_{r_2+r_3} 
                               \left( \frac{d-2}{2} \right)_{r_3} 
                              } 
z_1^{r_1} z_2^{r_2} z_3^{r_3}
\;, 
\label{H4}
\\
H_5^{(d) } & = & \sum_{r_1,r_2,r_3,r_4} 
                         \frac{\left( \frac{d-1}{2} \right)_{r_1+r_2+r_3+r_4} 
                               \left( \frac{d-2}{2} \right)_{r_2+r_3+r_4} 
                               \left( \frac{d-3}{2} \right)_{r_3+r_4} 
                               \left( \frac{d-4}{2} \right)_{r_4} 
                              }
                              {\left(\frac{d}{2} \right)_{r_1+r_2+r_3+r_4} 
                               \left( \frac{d-1}{2} \right)_{r_2+r_3+r_4} 
                               \left( \frac{d-2}{2} \right)_{r_3+r_4} 
                               \left( \frac{d-3}{2} \right)_{r_4} 
                              } 
z_1^{r_1} z_2^{r_2} z_3^{r_3} z_4^{r_4} 
\;. 
\label{H5}
\end{eqnarray}
In accordance with Eq.~(\ref{diff}), 
the order of differential equations of hypergeometric functions Eq.~(\ref{HYPERA}), 
increases with the number of external legs:
$$
\frac{P_N^j}{Q_N^j} = j \;,
$$
where the index $j$ is the same as the summation index $r_j$ and $j+1$ is equal 
to the number of external legs of the Feynman diagrams, cf., Eq.~(\ref{zero}).
To reduce the order of differential equations of the hypergeometric function
Eq.~(\ref{HYPERA}), we apply recursively the following transformation:
\begin{eqnarray}
\sum_{r=0}^\infty 
\frac{\Gamma\left(A \!+\! r \right)}
     {\Gamma\left(B \!+\! r \right)}
z^{r}
& = & 
\frac{\Gamma\left(A \right)}
     {\Gamma\left(B \right)}
{}_{2}F_1\left(\begin{array}{c|}
1, A  \\
   B  \end{array} ~z \right) 
\, = \,
\frac{1}{1-z}
\frac{\Gamma\left(A \right)}
     {\Gamma\left(B \right)}
{}_{2}F_1\left(\begin{array}{c|}
1, B \!-\! A \\
   B  \end{array} ~\frac{z}{z-1} \right) 
\nonumber \\  
& = & 
\frac{1}{1-z}
\frac{\Gamma\left(A \right)}
     {\Gamma\left(B \!-\! A \right)} 
\sum_{r=0}^\infty 
\left( \frac{z}{z-1} \right)^r
\frac{\Gamma\left(B \!-\! A \!+\! r \right)}
     {\Gamma\left(B \!+\! r \right)}
\;.
\label{rec}
\end{eqnarray}
Let us introduce new variables: 
\begin{equation}
x_i = - \frac{z_i}{1-z_i} \;, 
\quad 
z_i = - \frac{x_i}{1-x_i} \;, 
\quad 
1-z_i = \frac{1}{1-x_i} \;.
\label{xi-def}
\end{equation}
The recursive application of the linear-fractional transformation, Eq.~(\ref{rec}), 
to Eq.~(\ref{HYPERA}) gives rise to the following hypergeometric function:
\begin{eqnarray}
{\lefteqn{
\prod\limits_{k=1}^{N-1} (1-z_k) H_{N \geq 2}^{(d)} 
=  }}
\nonumber\\
&=& 
\sum_{r_1,r_2,r_3,r_4,\cdots,r_{N-1}=0}^\infty \,
\frac{\Gamma\left(\frac{d}{2}\right)
}
{\Gamma\left(\frac{d}{2} \!+\! r_1 \!+\! \cdots \!+\! r_{N-1} \right)
}
\prod\limits_{i=1}^{N-1} \,
x_i^{r_i} \,
\left[ 
\frac{\Gamma\left(\frac{i}{2} \!+\! r_1 \!+\! \cdots \!+\! r_{i} \right) 
     } 
     {\Gamma\left(\frac{i}{2} \!+\! r_1 \!+\! \cdots \!+\! r_{i-1} \right) 
     }
\right]
\, .
\qquad
\label{HYPERB}
\end{eqnarray}

For completeness, we present explicitly the hypergeometric terms defined by Eq.~(\ref{HYPERB})
for the first few values of $N=2,3,4,5$:~\footnote{For our discussion we drop all irrelevant 
factors, like $(1-z_i)^{\pm 1}$ and assume, wherever it does not cause
any problems, that
$
\Gamma  \left( \frac{d}{2} \pm k + \vec{r} \right) 
\equiv  \left( \frac{d}{2} \pm k  \right)_{\vec{r}}
$ 
with $k$ being integer.} 
\begin{eqnarray}
H_2^{(d)} & = & 
\sum_{r_1} \frac{ \left( \frac{1}{2} \right)_{r_1}}{\left(\frac{d}{2} \right)_{r_1}}  x_1^{r_1}   
\;,
\\ 
H_3^{(d)} & = & 
\sum_{r_1,r_2} 
\frac{\left( \frac{1}{2} \right)_{r_1} \left( \frac{d-2}{2}\right)_{r_2}}
     {            \left(\frac{d}{2} \right)_{r_1+r_2}}  
x_1^{r_1} z_2^{r_2} \;,
\label{H3a}
\\
& = & 
\sum_{r_1,r_2} 
\frac{\left( \frac{1}{2} \right)_{r_1} (1)_{r_1+r_2}}
     {(1)_{r_1}            \left(\frac{d}{2} \right)_{r_1+r_2}}  
x_1^{r_1} x_2^{r_2} 
\;,
\label{H3b}
\end{eqnarray}
%
\begin{eqnarray}
H_4^{(d)} & = & 
\sum_{r_1,r_2,r_3} \frac{\left( \frac{1}{2} \right)_{r_1}  
                                     \left(\frac{d-2}{2}\right)_{r_2+r_3}
                                     \left(\frac{d-3}{2}\right)_{r_3}
                        } 
                        { 
                              \left(\frac{d}{2} \right)_{r_1+r_2+r_3}
                              \left(\frac{d-2}{2}\right)_{r_3}
} 
x_1^{r_1} z_2^{r_2} z_3^{r_3}
\label{H4a}
\\ & = & 
\sum_{r_1,r_2,r_3} \frac{\left( \frac{1}{2} \right)_{r_1}  (1)_{r_1+r_2} 
                                     \left(\frac{d-3}{2} \right)_{r_3}} 
                              {(1)_{r_1} \left(\frac{d}{2} \right)_{r_1+r_2+r_3}} 
x_1^{r_1} x_2^{r_2} z_3^{r_3}
\label{H4b}
\\
& \equiv & 
\sum_{r_1,r_2,r_3} \frac{\left( \frac{1}{2} \right)_{r_1}  (1)_{r_1+r_2} 
                                     \left(\frac{3}{2} \right)_{r_1+r_2+r_3}
                        } 
                              { (1)_{r_1}
                                \left(\frac{3}{2} \right)_{r_1+r_2}
                                \left(\frac{d}{2} \right)_{r_1+r_2+r_3}
                          } 
x_1^{r_1} x_2^{r_2} x_3^{r_3}
\label{H4c}
\;,
\end{eqnarray}
where the $x_i$ are defined in Eq.~(\ref{xi-def}).
As follows from Eq.~(\ref{H3a}) and Eq.~(\ref{H3b}), the vertex diagrams 
are described by the Appell hypergeometric functions $F_3$ or $F_1$ \cite{tarasov-vertex}.
For the pentagon $(N=5)$, the hypergeometric function has the following form: 
\begin{eqnarray}
H_5^{(d)} & = & 
\sum_{r_1,r_2,r_3,r_4}^\infty 
\frac{
      \left( \frac{1}{2} \right)_{r_1} 
      \left(\frac{d-2}{2} \right)_{r_2 \!+\! r_3 \!+\! r_{4}} 
     }
     { 
      \left(\frac{d}{2} \right)_{r_1 \!+\! r_2 \!+\! r_{4}}
     }
\frac{\left(\frac{d-3}{2} \right)_{r_3 \!+\! r_4} }
     {\left(\frac{d-2}{2} \right)_{r_3 \!+\! r_4} }
\frac{\left(\frac{d-4}{2}\right)_{r_4} }
     {\left(\frac{d-3}{2}\right)_{r_4} }
x_1^{r_1}
z_2^{r_2} z_3^{r_3} z_4^{r_4} \;,
\label{H5a}
\\ 
& = & 
\sum_{r_1,r_2,r_3,r_4=0}^\infty 
\frac{\left( \frac{1}{2} \right)_{r_1} 
      \left(1 \right)_{r_1 \!+\! r_2} }
     {\left(1 \right)_{r_1} } 
\frac{\left(\frac{d-3}{2} \right)_{r_3 \!+\! r_4} }
     {\left(\frac{d}{2} \right)_{r_1 \!+\! r_2 \!+\! r_3 \!+\! r_4} }
\frac{\left(\frac{d-4}{2} \right)_{r_4} }
     {\left(\frac{d-3}{2} \right)_{r_4} }
x_1^{r_1}
x_2^{r_2}
z_3^{r_3}  
z_4^{r_4}  \;,
\label{H5b}
\\ & = & 
\sum_{r_1,r_2,r_3,r_4=0}^\infty 
\frac{\left( \frac{1}{2} \right)_{r_1} 
      \left(1 \right)_{r_1 \!+\! r_2}
     } 
     {\left(1 \right)_{r_1}
     }
\frac{
      \left(\frac{3}{2} \right)_{r_1 \!+\! r_2 \!+\! r_3}
     } 
     {
      \left(\frac{3}{2} \right)_{r_1 \!+\! r_2}
     }
\frac{\left(\frac{d-4}{2} \right)_{r_4} }
     {\left(\frac{d}{2} \right)_{r_1 \!+\! r_2 \!+\! r_3 \!+\! r_4} }
x_1^{r_1}
x_2^{r_2}
x_3^{r_3}
z_4^{r_4}  \;,
\label{H5c}
\\ & = & 
\sum_{r_1,r_2,r_3,r_4} \frac{\left( \frac{1}{2} \right)_{r_1} (1)_{r_1+r_2} 
                             \left(\frac{3}{2} \right)_{r_1+r_2+r_3}
                             \left(\frac{4}{2} \right)_{r_1+r_2+r_3+r_4}
                              } 
                              {(1)_{r_1}
                              \left(\frac{3}{2} \right)_{r_1+r_2}
                              \left(\frac{4}{2} \right)_{r_1+r_2+r_3}
                              \left(\frac{d}{2} \right)_{r_1+r_2+r_3+r_4}
                               } 
x_1^{r_1} x_2^{r_2} x_3^{r_3} x_4^{r_4} 
\label{H5d}
\;.
\end{eqnarray}
After the linear-fractional transformation, Eq.~(\ref{rec}), 
the order of the differential equation for the hypergeometric function related 
to the box diagram is reduced from three to two, see Eqs.~(\ref{H4a}) and (\ref{H4b}),~\cite{fjt}.
The pentagon,  Eq.~(\ref{H5b}), corresponds to a hypergeometric function 
satisfying a differential equation of order two~\footnote{ 
At present, a full classification of Horn-type 
hypergeometric functions of four variables does not exist~\cite{multiple}.}.
As follows from Eqs.~(\ref{HYPERA}) and (\ref{HYPERB}),  
the massive hexagon is expressible in terms of hypergeometric functions of 
five variables satisfying a differential equations of order three.
However, since the difference between parameters of hypergeometric functions, 
Eqs.~(\ref{HYPERA}) or (\ref{HYPERB}), are integer or half-integer, 
these functions possess extended symmetries with respect 
to non-linear transformations of their arguments \cite{tarasov-nonlinear} 
(multivariable generalizations of quadratic transformations related 
to Gauss hypergeometric functions)~\footnote{
All hypergeometric functions, defined by Eqs.~(\ref{H2})-(\ref{H5d}), 
belong to the class of multiple Gauss hypergeometric functions \cite{multiple}: 
the series representation can be written as infinite sum(s)  
with respect to the index of summation 
over the parameters of ${}_2F_1$ hypergeometric functions: 
in Eqs.~(\ref{H3b}), (\ref{H4c}), (\ref{H5d}) the Gauss hypergeometric functions enter 
via the last index of summation; 
in Eq.~(\ref{H4a}) via summation over $r_1$;
in Eq.~(\ref{H4b}) via summation over $r_2$ or $r_3$;
in Eq.~(\ref{H5a}) via summation over $r_1$ or $r_2$;
in Eq.~(\ref{H5b}) via summation over $r_2$ or $r_3$;
in Eq.~(\ref{H5c}) via summation over $r_3$.
Exploring the transformation properties of Gauss hypergeometric functions, 
see Eq.~(\ref{rec}) for an example,
the transformation of hypergeometric functions, Eqs.~(\ref{H2})-(\ref{H5d}), can be performed. 
}.
It is still an open question, whether or not is it possible, 
to reduce the order of differential equations with the help of non-linear transformations. 

\subsection{Off-shell massless case}

Let us consider an off-shell massless one-loop $N$-point diagram, where 
for some $i$, we have $\{p_i^2\} \neq 0$, cf., Eq.~(\ref{zero}).
In these kinematics, the three-point diagram (vertex) is not algebraically reducible to a simpler diagram.
The $I_{2}^{(d)}$ integral can be written (up to some irrelevant normalization) as
\begin{equation}
I_2^{(d)} = \frac{1}
                 {\Gamma\left( \frac{d-1}{2} \right)} \;,
\end{equation}
and the iterative solution of Eq.~(\ref{tarasov}) is 
\begin{eqnarray}
&& 
\left. 
H^{(d)}_{
N \geq 3
}
\right|_{\{p_i^2\} \neq 0}
= 
\sum_{r_1,r_2,r_3,\cdots,r_{N-2}=0}^\infty 
\frac{\Gamma\left(\frac{d-2}{2} \!+\! r_1 \!+\! r_2 \cdots \!+\! r_{N-2} \right)}
     {\Gamma\left(\frac{d-1}{2} \!+\! r_1 \!+\! r_2 \cdots \!+\! r_{N-2} \right)}
\nonumber \\ && 
\times 
\frac{\Gamma\left(\frac{d-3}{2} \!+\! r_2 \!+\! r_3 \cdots \!+\! r_{N-2} \right)}
     {\Gamma\left(\frac{d-2}{2} \!+\! r_2 \!+\! r_3 \cdots \!+\! r_{N-2} \right)}
\frac{\Gamma\left(\frac{d-4}{2} \!+\! r_3 \!+\! r_4 \cdots \!+\! r_{N-2} \right)}
     {\Gamma\left(\frac{d-3}{2} \!+\! r_3 \!+\! r_4 \cdots \!+\! r_{N-2} \right)}
\nonumber \\ && 
\times 
\frac{\Gamma\left(\frac{d-5}{2} \!+\! r_4 \cdots \!+\! r_{N-2} \right)}
     {\Gamma\left(\frac{d-4}{2} \!+\! r_4 \cdots \!+\! r_{N-2} \right)}
\cdots 
\frac{\Gamma\left(\frac{d-N+1}{2} \!+\! r_{N-2} \right)}
     {\Gamma\left(\frac{d-N+2}{2} \!+\! r_{N-2} \right)}
z_1^{r_1} z_2^{r_2} \cdots z_{N-2}^{r_{N-2}} \;.
\label{HYPER-OFF}
\end{eqnarray}
From Eqs.~(\ref{HYPERA}) and (\ref{HYPER-OFF}) we see, that the structure of hypergeometric 
functions related with off-shell massive and off-shell massless integrals is related as follows \cite{kt}:
\begin{equation}
\left. 
H_{N+1}^{(d)} 
\right|_{\{p_i^2\} \neq 0}
\sim 
H_N^{(d-1)} \;,
\label{subs}
\end{equation}
where $d$ is the dimension of space-time and $N$ denotes the number of external legs.
The symbol $\sim$ in Eq.~(\ref{subs}) indicates that this relation is valid 
for hypergeometric functions related with the corresponding Feynman diagram.
Eq.~(\ref{subs}) is also valid for hypergeometric functions, Eq.~(\ref{HYPERB}), 
after application of the linear-fractional transformation, Eq.~(\ref{rec}).

\section{Differential reduction of Horn-type hypergeometric functions of three variables}
\subsection{System of differential equations}
Let us consider the system of linear differential operators  of  second order $L_j$
for the hypergeometric functions $\omega(\vec{z})$:
\begin{eqnarray}
&&
L_i \omega(\vec{z}):
\quad
\theta_i^2 \omega(\vec{z}) =
\left[ 
\theta_i 
\sum_{j=1; \{j \neq i\} }^3 
P_{ij} \theta_j 
+ \sum_{m=1}^3  R_{im} \theta_m
+ S_i 
\right] \omega(\vec{z})
\;,
\quad
i = 1, 2, 3,
\label{canonical} 
\end{eqnarray}
where 
$\vec{z} = (z_1,z_2,z_3)$
with
$z_1,z_2,z_3$ being variables,
$\{P_{i,j},R_{i,ab},S_j\}$ are rational functions,
$\theta_j = z_j \partial_{z_j}$ for $j=1,2,3$,
and
$
\theta_{i_1\cdots i_k} = \theta_{i_i} \cdots \theta_{i_k}.
$
Taking the derivative, 
$\theta_k L_i \omega(\vec{z})$, 
we finally obtain from Eq.~(\ref{canonical}):
\begin{eqnarray}
\theta_k L_i \omega(\vec{z}):
&& 
\left[ 
\left( 
1
- 
P_{ik} P_{ki} 
\right) 
\theta_k \theta_i^2 
- 
\sum_{J=1; \{J \neq i  \neq k\} }^3 
\left( 
P_{iJ} 
+ 
P_{ik} P_{kJ} 
\right)
\theta_i \theta_k  \theta_J
\right] 
\omega(\vec{z})
\nonumber \\ && \hspace{5mm}
= 
\Biggl\{ 
\left[ 
P_{ik} R_{ki} P_{ik} 
+ R_{ik} P_{ki} 
+ R_{ii} 
+ 
P_{ik} R_{kk} 
+ \left( \theta_k P_{ik} \right)
\right]
\theta_{ik}
\nonumber \\ && \hspace{5mm}
+ 
\sum_{J=1; \{J \neq i \neq k\} }^3 
\left[ 
  P_{ik} R_{ki} P_{iJ} 
+ P_{ik} R_{kJ} 
+ \left( \theta_k P_{iJ} \right)
\right]
\theta_i \theta_J 
\nonumber \\ && \hspace{5mm}
+ 
\sum_{J=1; \{J \neq i \neq k\} }^3 
\left[ 
  R_{ik} P_{kJ} 
+ R_{iJ} 
\right]
\theta_k \theta_J
\nonumber \\ && \hspace{5mm}
+ 
\sum_{m=1}^3 
\left[ 
  P_{ik} R_{ki} R_{im} 
+ R_{ik} R_{km} 
+ \left( \theta_k R_{im} \right) 
\right]
\theta_m 
\nonumber \\ && \hspace{5mm}
+ P_{ik} S_k \theta_i 
+ S_i \theta_k 
+ P_{ik} R_{ki} S_i 
+ R_{ik} S_k 
+ \left( \theta_k S_i \right)  
\Biggr\} \omega(\vec{z})
\;, 
\quad 
i,k=1,2,3 \;. \quad 
\label{second}
\end{eqnarray}
For a function of three variables,  the sum $\sum_{J=1; \{J \neq i \neq k\} }^3$ 
can be replaced by the index $j$, where $j \neq i \neq k$.
The conditions of complete integrability are defined via the relations:
\begin{equation}
\theta_i \left[ \theta_j L_k  \right] \omega(\vec{z})
= 
\theta_j \left[ \theta_i L_k  \right] \omega(\vec{z})
\;, 
\quad 
i,j,k = 1,2,3. 
\label{integrability}
\end{equation}
The number of independent solutions of the system of differential equations, Eq.~(\ref{canonical}), of three variables 
is defined by coefficients on the l.h.s. of Eq.~(\ref{second}) and the validity of Eq.~(\ref{integrability}).
When the coefficients 
\begin{equation}
\left( 
1
- 
P_{ik} P_{ki} 
\right)
\;, 
\quad i,k=1,2,3 \;, 
\label{condition:1}
\end{equation}
and 
\begin{eqnarray}
\sum_{J=1; \{J \neq i \neq k\} }^3
\left( 
P_{iJ} 
+ 
P_{ik} P_{kJ} 
\right) \;,
\quad 
i,k=1,2,3 \;,
\label{condition:2}
\end{eqnarray}
are not equal to zero for all $i,J,k$, 
Eqs.~(\ref{canonical}) and (\ref{second}) can be reduced to the Pfaff system of 
eight independent differential equations: 
\begin{equation}
d \vec{f} = R \vec{f} \;, 
\label{Pfaff}
\end{equation}
where 
$
\vec{f} = \left( 
\omega(\vec{z}),
\theta_1 \omega(\vec{z}),
\theta_2\omega(\vec{z}),
\theta_3\omega(\vec{z}),
\theta_{12}\omega(\vec{z}),
\theta_{13}\omega(\vec{z}),
\theta_{23}\omega(\vec{z}),
\theta_{123}\omega(\vec{z})
\right).
$
When some of the coefficients in Eq.~(\ref{condition:1}) are zero, 
the coefficients in front of the terms $\theta_{123}\omega(\vec{z})$, defined 
by Eq.~(\ref{condition:2}), start to play a role.
For non-zero values of Eq.~(\ref{condition:2}), the terms $\theta_{123}\omega(\vec{z})$ can be excluded, and 
the rank of differential system is reduced to seven independent functions. 
When for some values of $i$ and $k$ both coefficients, defined by 
Eq.~(\ref{condition:1}) and Eq.~(\ref{condition:2}) are zero, 
a further simplification can be performed, so that the rank of system is reduced to six or 
to an even smaller number.

The locus of singularities $L_{ij}$ of the 
linear system of differential equations of second order of three variables defined by  Eq.~(\ref{canonical}) 
follows from singularities of higher rank differential operators on the l.h.s.
of Eq.~(\ref{canonical}) and Eq.~(\ref{second}):  
\begin{eqnarray}
&& \hspace*{-7mm}
L_{ij} = 
       \cup_{i=1}^3 \{z_i=0\} 
       \cup_{i,k=1}^3 \{P^{-1}_{ik}=0\} 
       \cup_{i,k=1}^3 \{\left(1 \!-\! P_{ik} P_{ki} \right)^{-1}=0\} 
       \cup_{i,j,k=1}^3 \{\left(P_{ij} \!+\! P_{ik} P_{kj}\right)^{-1}=0\} 
\;.
\nonumber \\ && \hspace*{-7mm}
\label{locus}
\end{eqnarray}
This result is valid only for the full (non-degenerate) case of Pfaff system including eight elements.

\subsection{Lauricella hypergeometric function $F_D$}
\subsubsection{General consideration}
Let us consider the $F_D^{(r)}$ functions of $r$ variables, defined around $z_i=0$ as 
\begin{eqnarray}
F^{(r)}_D(a;b_1, \cdots, b_k; c; z_1, \cdots, z_r)
= \sum_{m_1,\cdots, m_r=0}^\infty
\frac{(a)_{|\vec{m}|}}{(c)_{|\vec{m}|}}  \prod\limits_{j=1}^r(b_j)_{m_j}
\frac{z_1^{m_1}}{m_1!} \cdots  \frac{z_k^{m_r}}{m_r!} 
\;, 
\label{FD-definition}
\end{eqnarray}
For $r=1$ this functions coincides with the Gauss hypergeometric function, 
for $r=2$, it coincides with Appell function $F_1$ \cite{appell}.
As follows from the definition, Eq.~(\ref{FD-definition}), this function 
is symmetric with respect to the transformation
$$
b_i \Leftrightarrow b_j \;, \quad
z_i  \Leftrightarrow z_j  \;.
$$
Generally, 
$F_D$ functions and their properties have been analyzed in detail in many references~\cite{miller,book}. 
The differential operators for $F_D$ hypergeometric function,
Eq.~(\ref{diff}), are given by
\begin{eqnarray}
&& 
D_i F^{(r)}_D: 
\quad  
\partial_i \left(c-1 \!+\! \sum_{j=1}^r \theta_j \right) F_D^{(r)}  =  
           \left(a \!+\!   \sum_{j=1}^r \theta_j \right) \left(b_i+\theta_i \right) F_D^{(r)}  \;,
\nonumber \\ && 
\quad 
i = 1, \cdots, r.
\label{FD:diff}
\end{eqnarray}
where 
\begin{equation}
F_D 
\equiv 
F_D^{(r)}(a;b_1,\cdots,b_r ;c; z_1, \cdots, z_r) \;. 
\label{FD}
\end{equation}
They can be written in canonical form, cf. Eq.~(\ref{canonical}): 
\begin{eqnarray}
&& 
L_i F_D: 
\quad 
\theta_i^2 F_D = 
\left[ 
- \theta_i \sum_{j; j \neq i} \theta_j 
+ \frac{(a+b_i)z_i \!-\! (c\!-\!1)}{1\!-\!z_i} \theta_i 
+ \frac{b_i z_i}{1\!-\!z_i} \sum_{j; j \neq i} \theta_j
+ \frac{a b_i z_i}{1-z_i} 
\right] 
F_D \;, 
\nonumber \\ && 
\quad 
i = 1, \cdots, r.
\label{FD:canonical}
\end{eqnarray}
From these equations we have: 
\begin{eqnarray}
&&
P_{ij} = P_{ji} = -1 \;, 
\quad
S_{i} = \frac{ab_iz_i}{1-z_i} \equiv a P_i \;,
\nonumber \\ && 
R_{ii} = \frac{(a+b_i)z_i-(c-1)}{1-z_i} \equiv R_i \;, 
\quad 
R_{im} = \frac{b_i z_i }{1-z_i} \equiv P_i \;, \quad m \neq i \;.
\label{FD:PRS}
\end{eqnarray}
Upon substitution of these values for $P_{ij}, R_{ab}, S_i$ into Eq.~(\ref{second}), we obtain
\begin{eqnarray}
\hspace{-5mm}
\left( 
\left[ 
  P_k 
\!-\! P_i
\!+\! R_{i} 
\!-\! R_{k} 
\right]
\theta_i \theta_k
\!-\! \left[ 
R_{ki} R_{im} 
\!-\! R_{ik} R_{km} 
\right]
\sum_{m=1}^r \theta_m 
\!-\! S_k \theta_i 
\!+\! S_i \theta_k 
\!-\! P_k S_i 
\!+\! P_i S_k 
\right) F_D
= 0 
\;.
\label{FD:3}
\end{eqnarray}
Eq.~(\ref{FD:3}) can be simplified with the help of Eq.~(\ref{FD:PRS}) by 
taking into account that the sum of the last two terms in Eq.~(\ref{FD:3}),  
$
P_i S_k - P_k S_i,  
$
is equal to zero, and by splitting the sum over $m$ into $i,k$  and $j$, where $j \neq i \neq k$.
In this way, we get
\begin{eqnarray}
&&
\Biggl(
   R_{k} 
\!-\!  P_k 
\!-\! R_{i} 
\!+\! P_i
\Biggr)
\theta_i \theta_k
F_D
= 
\Biggl( 
P_k
\left[ 
P_i - R_i
- a 
\right]
\theta_i 
- P_i
\left[ 
  P_k 
- R_k 
- a 
\right]
\theta_k 
\Biggr) F_D
\;,
\label{FD:5}
\end{eqnarray}
where 
\begin{equation}
R_i - P_i
\equiv
R_{ii} 
- 
R_{ik} 
= 
\frac{az_i-(c-1)}{1-z_i} \;. 
\end{equation}
Eq.~(\ref{FD:5}) can be rewritten in a more familiar form, see \cite{book}: 
\begin{eqnarray}
\left[ 
(z_i - z_j) \theta_i \theta_j - b_j z_j \theta_i +  b_i z_i \theta_j  
\right] F_D 
= 0 
\;. 
\label{FD:diff:2A}
\end{eqnarray}
After factorization of $z_i,z_j$, Eq.~(\ref{FD:diff:2A}) can be expressed as follows,
\begin{equation}
\left[ 
\left( z_i-z_j \right) \partial_{ij}
+ b_i \partial_j 
- b_j \partial_i
\right] F_D = 0.
\label{FD:diff:2B}
\end{equation}
In this way, all second derivatives of an $F_D$ function are expressible in terms 
of the corresponding first derivatives and function, see Eqs.~(\ref{FD:canonical}) and (\ref{FD:diff:2A}).
As consequence, there are only $r+1$ linearly independent solutions of 
linear differential equations, Eq.~(\ref{FD:diff}).
The locus of the singularities $L_{ij}$ of an $F_D$ function is defined from the singularities 
of the differential equations, Eqs.~(\ref{FD:canonical}) and (\ref{FD:diff:2A}):
\begin{equation}
L_{ij} = \cup_{i=1}^r \{z_i=0\} \cup_{1=i < j=r} \{z_i - z_j = 0 \} \cup_{i=1}^r \{z_i=1\} \;.
\label{FD:locus}
\end{equation}
The Pfaff system for an $F_D$ hypergeometric function has the following form: 
$$
d \omega(\vec{z}) = \left( \sum_{i<j} A_{ij} d \log (z_i - z_j) \right) \omega(\vec{z}) \;, 
$$
where $\omega(\vec{z})  = \{ F_D, \theta_j F_D \}$
and the matrices $A_{ij}$ have been constructed explicitly in \cite{book}.

\subsubsection{Differential reduction of $F_D$}
The direct differential operators are the following:
\begin{eqnarray}
&& 
a
F_D^{(r)}(a+1;b_1,\cdots,b_r ;c; z_1, \cdots, z_r) 
= 
\left( 
a \!+\! \prod\limits_{i=1}^r \theta_i
\right)
F_D \;, 
%
%
%
\nonumber \\ && 
b_i 
F_D^{(r)}(a;b_1,\cdots,b_i+1, \cdots, b_r ;c; z_1, \cdots, z_r) 
= 
\left( 
b_i \!+\!  \theta_i
\right)
F_D \;, 
\nonumber \\ && 
(c \!-\! 1)
F_D^{(r)}(a;b_1,\cdots,b_r ;c-1; z_1, \cdots, z_r) 
= 
\left( 
c \!-\! 1 \!+\!  \prod\limits_{i=1}^r \theta_i
\right)
F_D \;, 
\label{direct:FD}
\end{eqnarray}
and $F_D$ is defined by Eq.~(\ref{FD}).
The inverse differential operators have been constructed in \cite{miller}:
\begin{eqnarray}
&& 
(c\!-\!a) F^{(r)}_D(a\!-\!1;b_1,\cdots,b_r ;c; z_1, \cdots, z_r) 
= 
\nonumber \\ && 
\hspace{35mm}
\left[
\sum_{j=1}^r (1-z_j) \theta_j \!-\! \sum_{j=1}^r b_j z_j \!+\! c \!-\! a 
\right]
F_D\;,
\label{inverse:FD:a}
\\ && 
(c \!-\! \sum_{j=1}^r b_j) F_D^{(r)}(a; b_1,\cdots, b_i\!-\!1, \cdots, b_r ;c; z_1, \cdots, z_r) 
= 
\nonumber \\ && 
\hspace{35mm}
\left[
z_i \sum_{j=1}^r (1\!-\!z_j) \partial_j \!-\! a z_i \!+\! c \!-\! \sum_{j=1}^r b_j 
\right]
F_D  \;,
\\ && 
(c \!-\! a)
(c \!-\! \sum_{j=1}^r b_j) F_D^{(r)}(a; b_1,\cdots, b_r ;c\!+\!1; z_1, \cdots, z_r) 
= 
\nonumber \\ && 
\hspace{35mm}
c
\left[
\sum_{j=1}^r (1-z_j) \partial_j \!+\! c \!-\! a \!-\! \sum_{j=1}^r b_j 
\right]
F_D  \;,
\nonumber \\ && 
\label{inverse:FD:c}
\end{eqnarray}
where 
$
F_D 
$
is defined by Eq.~(\ref{FD}).
In this case, the results of the differential reduction, Eq.~(\ref{reduction}), have the following form 
%
%
\begin{equation}
S(\vec{z}) F_D((a;\vec{b};c)+\vec{m}; \vec{z})
= 
S_0(\vec{z}) F_D(a;\vec{b};c; \vec{z})
+ 
\sum_{i=1}^r S_i(\vec{z}) \frac{\partial}{\partial z_i} F_D^{(r)}(a;\vec{b};c; \vec{z}) \;,
\label{reduction:FD}
\end{equation}
where $\vec{m}$ is a set of integers and $S$, $S_j$ are polynomials.

\subsection{Hypergeometric function $F_S$}
\label{FS:section}
\subsubsection{General consideration}
The Lauricella-Saran hypergeometric function of three variables $F_S$  \cite{Saran}
($F_7$ in the notation of \cite{Lauricella}) is defined around the point $z_1=z_2=z_3=0$ as follows
\begin{eqnarray}
F_S(a_1;a_2;b_1,b_2,b_3; c; z_1,z_2,z_3)
= \sum_{m_1,m_2,m_3=0}^\infty
\frac{(a_1)_{m_1} (a_2)_{m_2+m_3}}      
     {(c)_{m_1+m_2+m_3}} \prod\limits_{j=1}^3 (b_j)_{m_j} 
\frac{z_1^{m_1}  z_2^{m_2}  z_3^{m_3}}{m_1! m_2! m_3!} 
\;. 
\label{FS:series}
\end{eqnarray}
It is one of the $14$ functions of three variables of order two~\footnote{The complete 
set of programs for the differential reduction for other functions from the Lauricella-Srivastava list~\cite{multiple}
will be presented in separate publication.}, introduced by Lauricella \cite{Lauricella}.
In this case, the differential operators, Eq.~(\ref{diff}), are 
\begin{eqnarray}
D_1 F_S: && 
\quad 
\partial_1 \left(c \!-\! 1 \!+\! \sum_{j=1}^3 \theta_j \right) F_S =  
\left(a_1 \!+\! \theta_1 \right) \left(b_1 \!+\! \theta_1 \right) F_S \;,
\end{eqnarray}
\begin{eqnarray}
D_i F_S: && 
\quad 
\partial_i \left(c-1 \!+\! \sum_{j=1}^3 \theta_j \right) F_S =  
\left(a_2 \!+\! \theta_2 \!+\! \theta_3 \right) \left(b_i \!+\! \theta_i \right) F_S
\;, 
\quad 
i = 2,3
\;,
\end{eqnarray}
where 
\begin{equation}
F_S 
= 
F_S(a_1,a_2;b_1,b_2,b_3;c; z_1,z_2,z_3) \;.
\label{FS}
\end{equation}
The canonical form of these differential equations are the following:  
\begin{eqnarray}
L_1 F_S: &&  
\theta_1^2 F_S = 
\left[ 
- \frac{1}{1-z_1} \theta_1 \left(\theta_2 \!+\! \theta_3  \right)
+ \frac{(a_1+b_1)z_1 - (c-1)}{1-z_1} \theta_1 
+ \frac{a_1 b_1 z_1}{1-z_1} 
\right] F_S
\;, 
\qquad
\label{L1FS}
\\ 
L_2 F_S: &&  
\theta_2^2 F_S = 
\left[ 
- \theta_2 \theta_3
\!-\! \frac{1}{1\!-\!z_2} \theta_2 \theta_1 
\!+\! \frac{(a_2\!+\!b_2)z_2 \!-\! (c\!-\!1)}{1\!-\!z_2} \theta_2 
\!+\! \frac{b_2 z_2}{1\!-\!z_2} \theta_3
\!+\! \frac{a_2 b_2 z_2}{1\!-\!z_2} 
\right] F_S 
\;, 
\qquad
\label{L2FS}
\\ 
L_3 F_S: &&  
\theta_3^2 F_S = 
\left[ 
- \theta_3 \theta_2
\!-\! \frac{1}{1\!-\!z_3} \theta_3 \theta_1 
\!+\! \frac{(a_2\!+\!b_3)z_3 \!-\! (c\!-\!1)}{1\!-\!z_3} \theta_3 
\!+\! \frac{b_3 z_3}{1\!-\!z_3} \theta_2
\!+\! \frac{a_2 b_3 z_3}{1\!-\!z_3} 
\right] F_S
\;. 
\qquad
\label{L3FS}
\end{eqnarray}
These equations define the values of functions $P_{ij},R_{ab},S_i$ entering in Eq.~(\ref{canonical}):
\begin{eqnarray}
&& 
R_{12} =  R_{13} = R_{21} = R_{31} = 0 \;,
\quad 
P_{23} = P_{32} = -1 \;, 
\quad
\nonumber \\ && 
P_{12} = P_{13}  = - \frac{1}{1-z_1} \;, 
\quad 
P_{21}  = - \frac{1}{1-z_2} \;, 
\quad 
P_{31}  = - \frac{1}{1-z_3} \;, 
\quad  
\nonumber \\ && 
R_{11} = \frac{(a_1 \!+\! b_1)z_1 \!-\! (c-1)}{1-z_1} \;, 
\quad 
R_{ii} = \frac{(a_2 \!+\! b_i)z_i \!-\! (c-1)}{1-z_i} \;, i = 2,3 \;, 
\nonumber \\ && 
R_{23} = \frac{b_2 z_2}{1-z_2} \;, 
\quad 
R_{32} = \frac{b_3 z_3}{1-z_3} \;, 
\quad 
S_1 = \frac{a_1 b_1 z_1}{1-z_1} \;, 
\quad 
S_i = a_2 \frac{b_i z_i}{1-z_i} \;, i = 2,3 \;.
\end{eqnarray}
With the substitution of these values of $P_{ij}$ into Eq.~(\ref{second}) and, since $1-P_{23}P_{32} = 0$, 
we can express the third mixing derivatives of the hypergeometric function, $\theta_{123}\omega(\vec{z})$, 
via second derivatives of the hypergeometric function. 

The series representation of the hypergeometric function $F_S$ can be rewritten in the following form: 
\begin{eqnarray}
&& 
F_S(a_1,a_2,a_2;b_1,b_2,b_3; c; z_1,z_2,z_3)
\nonumber \\ && 
= 
\sum_{m_1=0}^\infty
\frac{(a_1)_{m_1}  (b_1)_{m_1}} 
     {(c)_{m_1}} \frac{z_1^{m_1} }{m_1!} 
F_1(a_2;b_2,b_3;c\!+\!m_1; z_2,z_3) 
\;, 
\label{FS:F1}
\end{eqnarray}
where $F_1(a;b_1,b_2;c;z_1,z_2)$ is the Appell function of two variables: $F_1 \equiv F_D^{(2)}.$
From this representation it is easy to get the following relation:
\begin{eqnarray}
\left[ (z_2-z_3) \theta_{23} \right] F_S = \left( b_3 z_3 \theta_2 - b_2 z_2 \theta_3 \right) F_S \;.
\label{FS23}
\end{eqnarray}
Eq.~(\ref{second}) allows us to express all higher derivatives of hypergeometric functions $F_S$
in terms of second derivatives only. In particular, 
\begin{eqnarray}
&& 
\theta_2 L_1 : 
\left( 
1 - \frac{1}{(1-z_1)(1-z_2)} 
\right)
\theta_{112} F_S
\nonumber \\ && 
= 
\Biggl \{ 
(P_{12} R_{22} \!+\! R_{11}) \theta_{12}
\!+\! P_{12}  R_{23} \theta_{13}
\!+\! P_{12} S_2 \theta_1 
\!+\! S_1 \theta_2
\Biggr \} F_S 
\nonumber \\ && 
= 
\Biggl \{ 
\left(
\frac{(a_1\!+\!b_1)z_1-(c\!-\!1)}{1\!-\!z_1} 
- 
\frac{(a_2\!+\!b_2)z_2-(c\!-\!1)}{(1\!-\!z_1)(1\!-\!z_2)} 
\right) \theta_{12}
\nonumber \\ && 
- \frac{b_2 z_2}{(1\!-\!z_1)(1\!-\!z_2)} \theta_{13}
\!-\! \frac{a_2 b_2 z_2}{(1\!-\!z_1)(1\!-\!z_2)} \theta_{1}
\!+\! \frac{a_1 b_1 z_1 }{1-z_1} \theta_2
\Biggr \} F_S 
\;, 
\label{FS112}
\end{eqnarray}
\begin{eqnarray}
&& 
\theta_3 L_1 : 
\left( 
1 - \frac{1}{(1-z_1)(1-z_3)} 
\right)
\theta_{113} F_S
\nonumber \\ && 
= 
\Biggl \{
(P_{13} R_{33} + R_{11}) \theta_{13}
+ P_{13}  R_{32} \theta_{12}
+ P_{13}S_3 \theta_1 
+ S_1 \theta_3
\Biggr\} F_S 
\nonumber \\ && 
= 
\Biggl \{ 
\left(
\frac{(a_1\!+\!b_1)z_1-(c\!-\!1)}{1\!-\!z_1} 
- 
\frac{(a_2\!+\!b_3)z_3-(c\!-\!1)}{(1\!-\!z_1)(1\!-\!z_3)} 
\right) \theta_{13}
\nonumber \\ && 
- \frac{b_3 z_3}{(1\!-\!z_1)(1\!-\!z_3)} \theta_{12}
\!-\! \frac{a_2 b_3 z_3}{(1\!-\!z_1)(1\!-\!z_3)} \theta_{1}
\!+\! \frac{a_1 b_1 z_1 }{1-z_1} \theta_3
\Biggr \} F_S \;,
\label{FS113}
\end{eqnarray}
\begin{eqnarray}
&& 
\theta_3 L_2 = - \theta_2 L_3 : 
\nonumber \\ && 
\frac{(z_3-z_2)}{(1-z_2)(1-z_3)} 
\theta_{123} F_S 
= 
\Biggl\{ 
P_{21}  R_{32} \theta_{12}
- 
P_{31}  R_{23} \theta_{13}
\nonumber \\ && 
- \left( R_{22} \!-\! R_{23} \!+\! R_{32} \!-\! R_{33} \right) \frac{1}{z_2-z_3} 
\left(b_3 z_3 \theta_2 \!-\!  b_2 z_2 \theta_3 \right) 
\nonumber \\ && 
- 
R_{23} \left(a_2 \!+\! R_{33} \!-\! R_{32} \right) \theta_{3}
+ 
R_{32} \left(a_2 \!+\! R_{22} \!-\! R_{23} \right) \theta_{2}
\Biggr\} F_S 
\nonumber \\ && 
= 
\Biggl\{ 
  \frac{b_2 z_2 }{(1-z_2) (1-z_3)} \theta_{13}
- \frac{b_3 z_3 }{(1-z_2) (1-z_3)} \theta_{12}
\Biggr\} F_S \;.
\label{FS123}
\end{eqnarray}
The last equation, Eq.~(\ref{FS123}), coincides with the derivative of the Eq.~(\ref{FS23}) with respect to  $z_1$.
The remaining two differential equations we can write in the following form: 
\begin{eqnarray}
&& 
\theta_1 L_2 : 
\left( 
1 - \frac{1}{(1-z_1)(1-z_2)} 
\right)
\theta_{122} F_S 
= 
- \left(1 - \frac{1}{(1-z_1)(1-z_2)}\right) \theta_{123} F_S
\nonumber \\ && 
\hspace{10mm}
+ 
\Biggl\{ 
\left( P_{21} R_{11} + R_{22} \right) \theta_{12}
+ R_{23} \theta_{13}
+ P_2  S_1 \theta_{2}
+ S_2 \theta_{1}
\Biggr\} F_S 
\nonumber \\ && 
= 
- \left(1 - \frac{1}{(1-z_1)(1-z_2)}\right) \theta_{123} F_S
\nonumber \\ && 
\hspace{10mm}
+ 
\Biggl\{ 
\left(
\frac{(a_2\!+\!b_2)z_2-(c\!-\!1)}{1\!-\!z_2} 
- 
\frac{(a_1\!+\!b_1)z_1-(c\!-\!1)}{(1\!-\!z_1)(1\!-\!z_2)} 
\right) \theta_{12}
\nonumber \\ && 
\hspace{10mm}
\!+\! \frac{b_2 z_2}{(1\!-\!z_2)} \theta_{13}
\!-\! \frac{a_1 b_1 z_1}{(1\!-\!z_1)(1\!-\!z_2)} \theta_{2}
\!+\! \frac{a_2 b_2 z_2 }{1-z_2} \theta_1
\Biggr\} F_S  \;, 
\label{FS122}
\end{eqnarray}
\begin{eqnarray}
&& 
\theta_1 L_3 : 
\left( 
1 - \frac{1}{(1-z_1)(1-z_3)} 
\right)
\theta_{133} F_S 
= 
- \left(1 - \frac{1}{(1-z_1)(1-z_3)}\right) \theta_{123} F_S
\nonumber \\ && 
\hspace{10mm}
+ 
\left\{ 
\left( P_{31} R_{11} \!+\! R_{33} \right) \theta_{13}
+ R_{32} \theta_{12}
+ P_{31}  S_1 \theta_{3}
+ S_3 \theta_{1}
\right\} F_S 
\nonumber \\ && 
= 
- \left(1 - \frac{1}{(1-z_1)(1-z_2)}\right) \theta_{123} F_S
\nonumber \\ && 
\hspace{10mm}
+ 
\Biggl\{ 
\left(
\frac{(a_2\!+\!b_3)z_3-(c\!-\!1)}{1\!-\!z_3} 
- 
\frac{(a_1\!+\!b_1)z_1-(c\!-\!1)}{(1\!-\!z_1)(1\!-\!z_3)} 
\right) \theta_{13}
\nonumber \\ && 
\hspace{10mm}
\!+\! \frac{b_3 z_3}{(1\!-\!z_3)} \theta_{12}
\!-\! \frac{a_1 b_1 z_1}{(1\!-\!z_1)(1\!-\!z_3)} \theta_{3}
\!+\! \frac{a_2 b_3 z_3 }{1-z_3} \theta_1
\Biggr\} F_S  \;, 
\label{FS133}
\end{eqnarray}
where the mixed derivative $\theta_{123} F_S$ is defined by
Eq.~(\ref{FS123}). 

\noindent
In this way, we have proven: \\
\indent
{\bf Theorem 1:}\\
The Lauricella-Saran hypergeometric function $F_S$ of three variables, Eq.~(\ref{FS:series}), 
has six linearly independent solutions around the points $z_1=z_2=z_3=0$. 

The locus of singularities $L_{ij}$ of the hypergeometric function $F_S$
follows from the singularities of the differential operators, 
Eqs.~(\ref{L1FS})-(\ref{L3FS}), (\ref{FS123})-(\ref{FS133}):
\begin{equation}
L_{ij} = \cup_{i=1}^3 \{z_i=0\} \cup_{i=1}^3 \{z_i=1\} \cup \{z_2 = z_3\} 
\cup_{i=2}^3 \{z_1+z_i=z_1 z_i\} \;.
\label{FS:locus}
\end{equation}

\subsubsection{Differential reduction of $F_S$}
The direct differential operators are the following:
\begin{eqnarray}
a_1
F_S(a_1+1,a_2;\vec{b};c; \vec{x}) 
&=& 
\left( 
a_1 \!+\! \theta_1
\right) F_S 
\;, 
\nonumber \\ 
a_2
F_S(a_1,a_2+1;\vec{b};c; \vec{x}) 
&=& 
\left( 
a_2 \!+\! \theta_2 \!+\! \theta_3
\right)
F_S \;, 
\nonumber \\ 
b_i 
F_S(a_1,a_2;\cdots, b_i+1,\cdots;c; \vec{x}) 
&=& 
\left( 
b_i \!+\!  \theta_i
\right)
F_S \;, 
\nonumber \\ 
(c \!-\! 1)
F_S(\vec{a};\vec{b};c-1; \vec{x}) 
&=& 
\left( 
c \!-\! 1 \!+\!  \prod\limits_{j=1}^3 \theta_j
\right)
F_S \;. 
\label{direct:FS}
\end{eqnarray}
and $F_S$ is defined by Eq.~(\ref{FS}).
The corresponding inverse differential operators we define 
for the parameters ${\bf X} \in \{a_1,a_2,b_1,b_2,b_3\}$ as follows:
%
%
\begin{eqnarray}
&& 
F_S({\bf X};c;\vec{z})
= 
\left[ 
A_{{\bf X},Fs} 
\!+\! 
B_{{\bf X},Fs} \theta_1 
\!+\! 
C_{{\bf X},Fs} \theta_2 
\!+\! 
D_{{\bf X},Fs} \theta_3
\!+\! 
E_{{\bf X},Fs} \theta_{12}
\!+\! 
F_{{\bf X},Fs} \theta_{13}
\right] 
F_S({\bf X}+1;c;\vec{z}) \;, 
\nonumber \\ 
\end{eqnarray}
and for the parameter $c$:
\begin{eqnarray}
&& 
F_S(a_1,a_2;b_1,b_2,b_3;c;\vec{z})
= 
\nonumber \\ && 
\left[ 
A_{c,Fs} 
\!+\! 
B_{c,Fs} \theta_1 
\!+\! 
C_{c,Fs} \theta_2 
\!+\! 
D_{c,Fs} \theta_3
\!+\! 
E_{c,Fs} \theta_{12}
\!+\! 
F_{c,Fs} \theta_{13}
\right] 
F_S(a_1,a_2;b_1,b_2,b_3;c-1;\vec{z})
\;. 
\qquad
\end{eqnarray}
The full list of inverse differential operators are the following: 
%
%
\begin{eqnarray}
A_{a_1,Fs}
&=&
\frac{a_1^2+a_1 (b_1 z_1 \!+\! D_1 \!+\! D_3 \!-\! 2 b_1)
 + a_2 (b_1 z_1 \!+\! D_2 \!-\! a_1)
 +(b_1 z_1\!-\!c\!+\!1) (D_2 \!-\! a_1)}
{D_0 D_2}
\;, 
\nonumber \\
B_{a_1,Fs}
&=&
\frac{(z_1-1)(a_2\!+\!D_2)}
     {D_0 D_2}
\;,
\quad
C_{a_1,Fs}
 =  
\frac{b_1 z_1 (z_2\!-\!1)}
     {z_2 D_0 D_2}
\;,
\quad 
D_{a_1,Fs}
 =  
\frac{b_1 z_1 (z_3\!-\!1)}
     {z_3 D_0 D_2}
\;,
\nonumber \\
E_{a_1,Fs}
& = & 
-\frac{z_1\!+\!z_2\!-\!z_1 z_2}
      {z_2 D_0 D_2}
\;, 
\quad
F_{a_1,Fs}
=
-\frac{z_1\!+\!z_3\!-\!z_1 z_3}
      {z_3 D_0 D_2}
\;,
\label{inverse:FS:a1}
\end{eqnarray}
\begin{eqnarray}
A_{a_2,Fs}
&=&
\frac{(b_2 z_2 \!+\! b_3 z_3 \!+\! D_1) (a_1 \!+\! D_1 ) \!-\! b_1 D_1 }
     {D_0 D_1}
\;, 
\quad
B_{a_2,Fs}
=
\frac{(z_1\!-\!1) (b_2 z_2\!+\!b_3 z_3)}{z_1 D_0 D_1}
\;, 
\nonumber \\ 
C_{a_2,Fs}
& = & 
\frac{(z_2 \!-\!1) (b_1\!+\!D_0)}{D_0 D_1}
\;, 
\quad
D_{a_2,Fs}
=
\frac{(z_3\!-\!1) (b_1\!+\!D_0)}{D_0 D_1}
\;, 
\nonumber \\
E_{a_2,Fs}
&=&
-\frac{z_1+z_2-z_1 z_3}{z_1 D_0 D_1}
\;, 
\quad
F_{a_2,Fs}
=
-\frac{z_1+z_3-z_1 z_3}{z_1 D_0 D_1}
\;, 
\end{eqnarray}
\begin{eqnarray}
A_{c,Fs}
&=&
- \frac{(c-1)} {D_0 D_1 D_2 D_3} 
\Biggl[
a_1 (a_2+D_3)(D_1 \!+\! D_3)
+ D_1 (D_2 \!+\! a_2 \!-\! a_1) D_3  
 \Biggr]
\;, 
\\
%
%
%
%
%
B_{c,Fs}
&=&
-\frac{(c-1) (z_1-1) \left(a_2 ( D_1 \!+\! D_2) \!+\! D_2 D_3 \right)}
      {z_1 D_0 D_1 D_2 D_3 }
\;, 
\nonumber \\
C_{c,Fs}
&=&
\frac{(c-1) (1-z_2) 
\left(a_1 (D_1 \!+\! D_2) \!+\! D_1 D_3 \right)}
      {z_2 D_0 D_1 D_2 D_3 }
\;, 
\nonumber \\
D_{c,Fs}
& = & 
\frac{(c-1) (1-z_3) 
\left(a_1 (D_1 \!+\! D_2) \!+\! D_1 D_3 \right)}
      {z_3 D_0 D_1 D_2 D_3}
\;, 
\nonumber \\
E_{c,Fs}
&=&
\frac{(c-1) (z_1 + z_2- z_1 z_2) (D_1 + D_2)}{z_1 z_2 D_0 D_1 D_2 D_3}
\;, 
\quad
F_{c,Fs}
=
\frac{(c-1) (z_1 + z_3- z_1 z_3) (D_1 + D_2)}
     {z_1 z_3 D_0 D_1 D_2 D_3} 
\;, 
\nonumber
\end{eqnarray}
%

%
%
\begin{eqnarray}
A_{b_1,Fs}
&=&
\frac{a_2 (a_1 z_1+D_3)+(a_1 z_1+D_1-a_2) D_3}
     {D_1 D_3}
\;, 
\quad 
B_{b_1,Fs}
=
\frac{(z_1-1) (a_2+D_3)}{D_1 D_3}
\;,
\quad
\nonumber \\
C_{b_1,Fs}
& = & 
\frac{a_1 z_1 (z_2-1)}{z_2 D_1 D_3}
\;, 
\quad
D_{b_1,Fs}
=
\frac{a_1 z_1 (z_3-1)}{z_3 D_1 D_3}
\;, 
\nonumber \\
E_{b_1,Fs}
& = & 
- \frac{z_1+z_2-z_1 z_2}{z_2 D_1 D_3}
\;, 
\quad
F_{b_1,Fs}
=
- \frac{z_1+z_3-z_1 z_3}{z_3 D_1 D_3}
\;,
\end{eqnarray}
\\
%
%
\begin{eqnarray}
A_{b_2,Fs}
&=&
\frac{a_1 (a_2 z_2+D_3)+D_3 (a_2 z_2+D_3-b_1)}{D_2 D_3}
\;, 
\quad
B_{b_2,Fs}
=
\frac{a_2 (z_1-1) z_2}{z_1 D_2 D_3}
\;, 
\nonumber \\
C_{b_2,Fs}
&=&
\frac{(z_2-1) (a_1+D_3)}{D_2 D_3}
\;, 
\quad
D_{b_2,Fs}
=
\frac{z_2 (z_3-1) (a_1+D_3)}{z_3 D_2 D_3}
\;, 
\nonumber \\
E_{b_2,Fs}
&=&
- \frac{z_1 + z_2 - z_1 z_2}{z_1 D_2 D_3}
\;, 
\quad
F_{b_2,Fs}
=
- \frac{z_2 (z_1 + z_3 -z_1 z_3)}{z_1 z_3 D_2 D_3}
\;.
\end{eqnarray}
\\
%
%
\begin{eqnarray}
A_{b_3,Fs}
&=&
\frac{a_1 (a_2 z_3+D_3)+D_3 (a_2 z_3+D_3-b_1)}
     {D_2 D_3}
\;, 
\quad
B_{b_3,Fs}
=
\frac{a_2 (z_1-1) z_3}
     {z_1 D_2 D_3}
\;, 
\nonumber \\
C_{b_3,Fs}
&=&
\frac{(z_2-1) z_3 (a_1+D_3)}{z_2 D_2 D_3}
\;, 
\quad
D_{b_3,Fs}
=
\frac{(z_3-1) (a_1+D_3)}{D_2 D_3}
\;, 
\nonumber \\
E_{b_3,Fs}
&=&
- \frac{z_3 (z_1 + z_2 - z_1 z_2)}{z_1 z_2 D_2 D_3}
\;, 
\quad
F_{b_3,Fs}
=
-\frac{z_1+z_3-z_1 z_3}{z_1 D_2 D_3 }
\;, 
\label{inverse:FS:b3}
\end{eqnarray}
where
\begin{eqnarray}
D_0 & = &  a_1 + a_2 - (c-1) \;,
\\
D_1 & = & a_2 + b_1 - (c-1) \;, 
\\
D_2 & = & a_1 + b_2 + b_3 - (c-1) \;, 
\\ 
D_3 & = & b_1 + b_2 + b_3 - (c-1) \;,
\end{eqnarray}
and 
\begin{equation}
D_1 + D_2 = D_0 + D_3 \;.
\end{equation}
The results of the differential reduction, Eq.~(\ref{reduction}), have the following form in this case:
\begin{eqnarray}
{\lefteqn{
S(\vec{z}) F_S((\vec{a};\vec{b};c)+\vec{m}; \vec{z})
= }}
\nonumber \\ &&
S_0(\vec{z}) F_S(\vec{a};\vec{b};c; \vec{z})
+ 
\sum_{i=1}^3 S_i(\vec{z}) \frac{\partial}{\partial z_i} F_S(\vec{a};\vec{b};c; \vec{z}) 
+ 
\sum_{j=2}^3 S_{1j}(\vec{z}) \frac{\partial^2}{\partial z_1 \partial z_j} F_S(\vec{a};\vec{b};c; \vec{z}) 
\;,
\label{reduction:FS}
\end{eqnarray}
where $\vec{m}$ is a set of integers, $S$, $S_j$ and $S_{ij}$ are polynomials.

\subsection{Exceptional values of parameters: $F_D$ and $F_S$}
\label{EXCEP}
It was pointed out in \cite{hyperdire:1}, that the subset of parameters for which 
the results of the differential reduction,
Eqs.~(\ref{reduction:FD}) and (\ref{reduction:FS}), have simpler forms,
can be defined from the conditions
\begin{itemize}
\item[(i)] that the hypergeometric function entering the l.h.s. of
  Eqs.~(\ref{inverse:FD:a})--(\ref{inverse:FD:c}),
  (\ref{inverse:FS:a1})--(\ref{inverse:FS:b3}), is expressible in terms 
  of simpler hypergeometric functions 
  ($F_D^{(r-1)}$ for $F_D^{(r)}$ and ${}_2F_1, F_1$ or $F_3$ for $F_S$ hypergeometric function);
\item[(ii)] that some of the coefficients entering the inverse differential relations are equal to zero (infinity).
\end{itemize}
For the hypergeometric functions $F_D (\equiv F_D^{(r)})$ and $F_S$, the exceptional sets of parameters are listed in Table~\ref{tab:1}.
\begin{table}
$$
\begin{tabular}[ht!]{|c|c|}
\hline
$F_D^{(r)}$ & $\{a, b_j, c\!-\!a, c\!-\sum_{j=1}^r\!b_j \} \in \mathbb{Z}$\\
\hline
$F_S$ & $\{ a_1, a_2, b_j, 
            c\!-\!a_1\!-\!a_2, 
            c\!-\!b_1\!-\!b_2\!-\!b_3, 
            a_1\!+\!b_2\!+\!b_3\!-\!c, 
            a_2\!+\!b_1\!-\!c \}\in \mathbb{Z}$ \\
\hline
\end{tabular}
$$
\vspace*{-3mm}
\caption{\small
Exceptional set of parameters for the hypergeometric functions $F_D^{(r)}$ and $F_S$.}
\label{tab:1}
\end{table}
%

\section{Mathematica based program for the 
 differential reduction of $F_D$ and $F_S$ hypergeometric functions}
In this section, we will present the {\tt Mathematica} based
programs {\bf FdFunction}  and {\bf FsFunction} 
for the differential reduction of Horn-type hypergeometric functions 
$F_D$ of $r$ variables and $F_S$ of three variables~\footnote{The programs
  have been tested for {\tt Mathematica} version $8.0$.}.
In particular, in the application to Lauricella functions $F_D$, the reduction algorithm, Eq.~(\ref{reduction}), 
has the following form:
\begin{eqnarray}
R(x,y) F_D^{(r)}(a\!+\!m_a;\vec{b}\!+\!\vec{m_b};c\!+\!m_c; \vec{z})
=
\left[
P_0(\vec{z})
\!+\! P_1(\vec{z}) \theta_{z_1}
\!+\! \dots
\!+\! P_r(\vec{z}) \theta_{z_r}
\right] F_D^{(r)}(a;\vec{b};c; \vec{z}) \;,
\label{reductionfd}
\end{eqnarray}
where $\vec{m_b}$, $m_a$, $m_c$ are sets of integers and $\vec{b}$ , $a$, $c$ denote the set of parameters. 
$R, P_i$ are some polynomial and $\theta_{z_i} = z_i \partial_{z_i} $.
The differential reduction algorithm in application to the Lauricella-Saran function $F_S$ is:
\begin{eqnarray}
{\lefteqn{
R(\vec{z} )F_S(\vec{a}+\vec{m}_a;\vec{b}+\vec{m}_b;c+m_c; \vec{z})
= }}
\nonumber \\ &&  
\left[
P_0(\vec{z})
+ P_1(\vec{z}) \theta_{z_1}
+ P_2(\vec{z}) \theta_{z_2}
+ P_3(\vec{z}) \theta_{z_3}
+ P_{12}(\vec{z}) \theta_{z_1}\theta_{z_2}
+ P_{13}(\vec{z}) \theta_{z_1}\theta_{z_3}
\right] F_S(\vec{a};\vec{b};c; \vec{z}) \;,
\qquad
\label{reductionfs}
\end{eqnarray}
where, again, 
$\vec{m_a}$, $\vec{m}_b$, $m_c$ denote sets of integers, $\vec{a}$ , $\vec{b}$, $c$ sets of parameters,
and $R, \{P_j\}, \{P_{ij}\}$ some polynomials.
  
The program is freely available from \cite{bytev:hyper} subject to the license conditions specified.
The current version, $1.0$, deals with non-exceptional values of a parameters only.

\subsection{Package FdFunction}
The package can be loaded in the standard way:
$$
<< \mathrm{"FdFunction.m"}
$$
and it includes the following basic routines:
\begin{eqnarray}
\label{FdDef}
{\bf FdIndexChange}[\mbox{changingVector}, \mbox{parameterVector}],
\end{eqnarray}
and
\begin{equation}
{\bf FdDiffSeries}[\dots] \;,
\\
{\bf FdSeries [\dots] }
\end{equation}
The list $"{\rm changingVector}"$ in Eq.~(\ref{FdDef}) provides the set of integers 
by which the values of parameters of the Lauricella function $F_D$ are to be changed, i.e., 
the vector $m_a,\{ \vec m_b\},m_c$ in Eq.~(\ref{reductionfd}).
The set of initial parameters  of $F_D$ function are defined in the list $"{\rm parameterVector}"$
corresponding to  the vector $a+m_a;\vec{b}+\vec{m}_b,c+m_c$ 
and arguments $\vec z $ in the l.h.s. of Eq.~(\ref{reductionfd}).

The  structure of the output of ${\bf FdIndexChange}[]$ is the following:
\begin{eqnarray}
\{ \{A_1,A_2,\dots, A_{r+1} \},\{ {\rm parameterVectorNew} \}\},
\end{eqnarray}
where
\begin{itemize}
\item[(i)]
$"{\rm parameterVectorNew}"$ is the set of new parameters of $F_D^{(r)}$ hypergeometric function;
\item[(ii)]
$A_1,A_2,\dots, A_{r+1} $ are the rational functions corresponding to
the ratios of $P_0/R, P_1/R$, $P_2/R$ $\dots$ of functions entering in Eq.~(\ref{reductionfd}).
\end{itemize}

The functions ${\bf FdDiffSeries}[]$ and ${\bf FdSeries}[]$ are 
designed for the numerical evaluation of $F_D$ hypergeometric functions. 
They return the Taylor series of $F_D$ in its derivatives, respectively:
\begin{eqnarray}
&&
{\bf FdDiffSeries}[\mbox{numberOfvariable}, \mbox{vectorInit}, \mbox{numbSer}] \;,
\label{FdDiffSeries}
\\ &&
{\bf FdSeries [ \mbox{vectorInit}, \mbox{numbSer}] } \;, 
\label{FdSeries}
\end{eqnarray}
where
\begin{itemize}
\item[(i)]
 $"\mbox{numberOfvariable}"$ is the list of variable numbers for differentiation;  
\item[(ii)]
 $"\mbox{vectorInit}"$ is the set of Fd parameters;  
\item[(iii)]
 $"\mbox{numbSer}"$ is the number of terms in Taylor expansion.
\end{itemize}

Let us present a number of examples for the usage\footnote{
All functions in the package {\bf HYPERDIRE} generate output without additional simplification.
This is done for the maximum efficiency of the algorithm.
To bring the output into  a simpler form, we recommend to use in addition the command {\bf Simplify}. 
All examples considered here have been treated with the command {\bf Simplify[\dots]}.
subsequent to the call of {\bf HYPERDIRE}.
}

\noindent
{\bf Example 1.}
Differential reduction of the hypergeometric function $F_D^{(2)}$ of two variables~\footnote{
When $r=2$, the Lauricella function $F_D$ coincides with the Appell function $F_1$ and 
the package {\bf AppellF1F4} \cite{hyperdire:1} can be used for the differential reduction.}.\\ 
\\
{\bf FdIndexChange[}\{$-1$,\{$1$,$0$\},$1$\},\,\{$a$,\{$b_1$,$b_2$\},$c$,\{$z_1$,$z_2$\}\}{\bf ]}
\\
\\
\begin{eqnarray}
&& 
\Biggl\{ \Biggl\{\frac{c (-1 + z_2) - b_2 z_2 + z_1 (-1 + a + b_1 (-1 + z_2) + z_2 - a z_2 + b_2 z_2)}{c (-1 + z_2)},
\nonumber \\ && 
\frac{1 - z_2 - b_2 z_2 + z_1 (-1 + b_1 (-1 + z_2) + z_2 + b_2 z_2) +
 a (-1 + z_1 + z_2 - z_1 z_2)}{(-1 + a) c (-1 + z_2)},
\nonumber \\ &&  
 \frac{1 - a (1 + z_1 (-2 + z_2)) - b_2 z_2 + c z_2 +
 z_1 (-2 - c + b_1 (-1 + z_2) + z_2 + b_2 z_2)}{(-1 + a) c (-1 + z_2)},
\nonumber \\ &&  
 \{-1 + a, \{1 + b_1, b_2\}, 1 + c, \{z_1, z_2\} \Biggr\} \Biggr\}
\end{eqnarray}
\\
\\
In an explicit form:
\begin{eqnarray}
&&
F_D^{(2)}(a;b_1,b_2,c;z_1,z_2)
=
\nonumber \\ &&
\Biggl[
  \frac{c (-1 + z_2) - b_2 z_2 + z_1 (-1 + a + b_1 (-1 + z_2) + z_2 - a z_2 + b_2 z_2)}{c (-1 + z_2)}
\nonumber \\ &&
+\frac{1 - z_2 - b_2 z_2 + z_1 (-1 + b_1 (-1 + z_2) + z_2 + b_2 z_2) +
 a (-1 + z_1 + z_2 - z_1 z_2)}{(-1 + a) c (-1 + z_2)} \theta_1
 \nonumber \\ &&
+ \frac{1 - a (1 + z_1 (-2 + z_2)) - b_2 z_2 + c z_2 +
 z_1 (-2 - c + b_1 (-1 + z_2) + z_2 + b_2 z_2)}{(-1 + a) c (-1 + z_2)}\theta_2
\Biggr]
\nonumber \\ &&
\times
F_D^{(2)}(a-1;b_1+1,b_2,c+1; z_1,z_2).
\end{eqnarray}
\\

\noindent
{\bf Example 2.}
Reduction of the hypergeometric function $F_D^{(3)}$ of three variables. \\
\\
{\bf FdIndexChange[}\{$-1$,\{$1$,$-1$,$0$\},$0$\},\,\{$a$,\{$b_1$,$b_2$,$b_3$\},$c$,\{$z_1$,$z_2$,$z_3$\}\}{\bf ]}
\\
\\
$
\{\{
 \frac{z_1-1}{z_2-1} , \frac{z_1-1}{(a-1) \left(z_2-1\right)} , \frac{z_1 \left(a-c+\left(b_2-1\right)
   z_2\right)-\left(a-c+b_2-1\right) z_2}{(a-1) \left(b_2-1\right) \left(z_2-1\right) z_2} , \frac{z_1-1}{(a-1)
   \left(z_2-1\right)}\}, 
   \\
\{ a-1 , \left\{b_1+1,b_2-1,b_3\right\} , c , \left\{z_1,z_2,z_3\right\}
\}\}
$
\\
\\
This has the explicit form:
\begin{eqnarray}
&&
F_D^{(3)}(a;b_1,b_2,b_3;c;z_1,z_2,z_3)
=
\nonumber \\ &&
\Biggl[
   \frac{z_1-1}{z_2-1}+ \frac{z_1-1}{(a-1) \left(z_2-1\right)}\theta_1
   +\frac{z_1 \left(a-c+\left(b_2-1\right)
      z_2\right)-\left(a-c+b_2-1\right) z_2}{(a-1) \left(b_2-1\right) \left(z_2-1\right) z_2}\theta_2
\nonumber \\ &&
+\frac{z_1-1}{(a-1)
   \left(z_2-1\right)}\theta_3
\Biggr]
F_D^{(3)}(a-1;b_1+1,b_2-1,b_3;c; z_1,z_2,z_3).
\end{eqnarray}

\noindent
{\bf Example 3.}
Reduction of hypergeometric function $F_D^{(5)}$ of five variables. \\
\\
{\bf FdIndexChange[}\{$-1$,\{$0$,$1$,$0$,$0$,$-1$\},$0$\},\,\{$a$,\{$b_1$,$b_2$,$b_3$,$b_4$,$b_5$\},$c$,\{$z_1$,$z_2$,$z_3$,$z_4$,$z_5$\}\}{\bf ]}
\\
\\
$
\{\{
\frac{z_2-1}{z_5-1},
\frac{z_2-1}{(a-1) \left(z_5-1\right)},
\frac{z_2-1}{(a-1) \left(z_5-1\right)},
\frac{z_2-1}{(a-1)   \left(z_5-1\right)},
\frac{z_2-1}{(a-1) \left(z_5-1\right)},
\frac{z_2 \left(a+\left(b_5-1\right) z_5-c\right)-z_5
   \left(a+b_5-c-1\right)}{(a-1) \left(b_5-1\right) \left(z_5-1\right)
   z_5}\},
   \\ 
   \left\{a-1,\left\{b_1,b_2+1,b_3,b_4,b_5-1\right\},c,\left\{z_1,z_2,z_3,z_4,z_5\right\}\right\}\}
$
\\
\\
This has the explicit form:
\begin{eqnarray}
&&
F_D^{(5)}(a;b_1,b_2,b_3,b_4,b_5;c;z_1,z_2,z_3,z_4,z_5)
=
\nonumber \\ &&
\Biggl[
   \frac{z_2-1}{z_5-1}
   +\frac{z_2-1}{(a-1) \left(z_5-1\right)}\theta_1
   +\frac{z_2-1}{(a-1) \left(z_5-1\right)}\theta_2
   +\frac{z_2-1}{(a-1)   \left(z_5-1\right)}\theta_4
   \nonumber \\ &&
   +\frac{z_2 \left(a\!+\!\left(b_5\!-\!1\right) z_5\!-\!c\right)
        \!-\!z_5 \left(a\!+\!b_5\!-\!c\!-\!1\right)}{(a\!-\!1) \left(b_5\!-\!1\right) \left(z_5\!-\!1\right)
      z_5}\theta_5
\Biggr]
F_D^{(5)}(a\!-\!1;b_1\!+\!1,b_2\!-\!1,b_3;c;z_1,z_2,z_3,z_4,z_5).
\nonumber
\end{eqnarray}
\\

The hypergeometric function $F_D$ is not built into the current version of {\tt Mathematica}. 
The series representation of the hypergeometric function $F_D^{(r)}$, Eq.~(\ref{FD-definition}), 
is implemented in our package. 
The functions  ${\bf FdDiffSeries}[]$ and ${\bf FdSeries}[]$ allow 
to make numerical cross-checks of the results of the differential reduction. 
The corresponding examples for using these functions are all collected in the
file {\tt example-FdFunction.m}, 
which is available in \cite{bytev:hyper}.
%
%
%
%
%
%
\subsection{Package {\bf FsFunction} }
Again, the program can be loaded in a standard way:
$$
<< \mathrm{"FsFunction.m"}
$$
and its structure and output are similar to the {\bf FdFunction} package.
The package {\bf FsFunction} includes the following basic routines:
\begin{eqnarray}
\label{FsDef}
{\bf FsIndexChange}[\mbox{changingVector}, \mbox{parameterVector}],
\end{eqnarray}
Here, again, $"{\rm changingVector}"$ is the list of integers by which the values of 
the parameters of the function $F_S$ are to be changed, i.e., 
the vectors $\vec{m}_a, \vec m_b ,m_c$ in Eq.~(\ref{reductionfs}), while 
the set of initial parameters  of the function $F_S$ is defined in the list $"{\rm parameterVector}"$
corresponding to the vector $\vec{a}\!+\!\vec{m}_a;\vec{b}\!+\!\vec{m}_b,c\!+\!m_c$ 
and the arguments $\vec{z} $ in the l.h.s. of Eq.~(\ref{reductionfs}).

The  structure of the output of ${\bf FsIndexChange}[]$ is the following:
\begin{eqnarray}
\{ A,B,C,D,E,F \},\{ {\rm parameterVectorNew} \}\},
\end{eqnarray}
where
\begin{itemize}
\item[(i)]
$"{\rm parameterVectorNew}"$ is the set of new parameters of the function $F_S$;
\item[(ii)]
$A,B,C,D,E,F$ are the rational functions corresponding to
the ratios of $P_0/R$ $P_1/R$, $P_2/R$, $P_3/R$, $P_{12}/R$ and $P_{13}/R$ 
entering  Eq.~(\ref{reductionfs}).
\end{itemize}

\noindent
{\bf Example 4}: Reduction of $F_S$.\\
\\
{\bf FsIndexChange[}\{$1$,$-1$,$0$,$0$,$0$,$0$\},\,\{$a_1$,$a_2$,$b_1$,$b_2$,$b_3$,$c$,$z_1$,$z_2$,$z_3$\}{\bf ]}
\\
\\
\begin{eqnarray}
&& 
\Biggl\{ \Biggl\{
  \frac{a_1+b_1 z_1+b_2+b_3-c+1}{a_1+b_2+b_3-c+1},
 -\frac{1-z_1}{a_1+b_2+b_3-c+1},
\nonumber\\ && \hspace{5mm}
 -\frac{z_2\left(-a_1-b_2-b_3+c-1\right)-b_1 z_1 \left(z_2-1\right)}
       {\left(a_2-1\right) z_2\left(a_1+b_2+b_3-c+1\right)},
\nonumber\\ && \hspace{5mm}
   -\frac{z_3 \left(-a_1-b_2-b_3+c-1\right)-b_1 z_1
   \left(z_3-1\right)}{\left(a_2-1\right) z_3 \left(a_1+b_2+b_3-c+1\right)},
\nonumber\\ && \hspace{5mm}
   -\frac{z_2-z_1\left(z_2-1\right)}{\left(a_2-1\right) z_2 \left(a_1+b_2+b_3-c+1\right)},
   -\frac{z_3-z_1\left(z_3-1\right)}{\left(a_2-1\right) z_3
   \left(a_1+b_2+b_3-c+1\right)} \Biggr\},
\nonumber\\ && 
   \Biggl\{a_1+1,a_2-1,b_1,b_2,b_3,c,z_1,z_2,z_3 \Biggr\} \Biggr\}
\end{eqnarray}
\\
\\
This has the explicit form:
\begin{eqnarray}
&&
F_S(a_1,a_2;b_1,b_2,b_3;c;z_1,z_2,z_3)
=
\nonumber \\ &&
\Biggl[
 \frac{a_1+b_1 z_1+b_2+b_3-c+1}{a_1+b_2+b_3-c+1}
-\frac{1-z_1}{a_1+b_2+b_3-c+1} \theta_1
 \nonumber \\ &&
-\frac{z_2\left(-a_1-b_2-b_3+c-1\right)-b_1 z_1 \left(z_2-1\right)}{\left(a_2-1\right) z_2
   \left(a_1+b_2+b_3-c+1\right)}\theta_2
 \nonumber \\ &&
    -\frac{z_3 \left(-a_1-b_2-b_3+c-1\right)-b_1 z_1
      \left(z_3-1\right)}{\left(a_2-1\right) z_3 \left(a_1+b_2+b_3-c+1\right)}\theta_3
 \nonumber \\ &&
 -\frac{z_2-z_1\left(z_2-1\right)}{\left(a_2-1\right) z_2 \left(a_1+b_2+b_3-c+1\right)}\theta_1\theta_2
  -\frac{z_3-z_1\left(z_3-1\right)}{\left(a_2-1\right) z_3
    \left(a_1+b_2+b_3-c+1\right)}\theta_1\theta_3
\Biggr]
\nonumber \\ &&
\times
F_S(a_1+1,a_2-1;b_1,b_2,b_3;c;z_1,z_2,z_3).
\end{eqnarray}

\noindent 
{\bf Example 5}: Reduction of $F_S$ \\
\\
{\bf FsIndexChange[}\{$1$,$0$,$0$,$0$,$1$,$2$\},\,\{$a_1$,$a_2$,$b_1$,$b_2$,$b_3$,$c$,$z_1$,$z_2$,$z_3$\}{\bf ]}
\\
\\
$
\{\{
  -\frac{b_1 z_1 \left(z_2-1\right) \left(-a_2 z_3-a_1+c\right)-z_2 \left(a_2 \left(z_3
   \left(b_3-c\right)+b_2 \left(z_3-1\right)\right)+c (c+1)\right)+a_2 b_3 z_3-a_2 c z_3+c^2+c}{c (c+1)
   \left(z_2-1\right)},
   \frac{z_1 \left(a_2 z_3+a_1-c\right)-a_2 z_3+c}{c (c+1)},
   \nonumber \\
   -\frac{z_2 \left(a_2-b_2 z_3-b_3
   z_3+b_2+c z_3-2 c-1\right)-a_2 z_3-b_1 z_1 \left(z_2-1\right) \left(z_3-1\right)+b_3 z_3+c+z_3}{c (c+1)\left(z_2-1\right)},
      \nonumber \\
   \frac{z_2 \left(z_3 \left(b_3-c\right)+b_2 \left(z_3-1\right)+c\right)+b_1 z_1
   \left(z_2-1\right) \left(z_3-1\right)-b_3 z_3+c z_3-c}{c (c+1) \left(z_2-1\right)},
   \frac{\left(z_1
   \left(z_2-1\right)-z_2\right) \left(z_3-1\right)}{c (c+1) \left(z_2-1\right)},
   -\frac{-z_3z_1+z_1+z_3}{c^2+c}\},
      \nonumber \\
   \left\{a_1+1,a_2,b_1,b_2,b_3+1,c+2,z_1,z_2,z_3\right\}\}
$
\\
\\
This has the explicit form:
\begin{eqnarray}
&&
F_S(a_1,a_2;b_1,b_2,b_3;c;z_1,z_2,z_3)
=
\nonumber \\ &&
\Biggl[
 -\frac{b_1 z_1 \left(z_2-1\right) \left(-a_2 z_3-a_1+c\right)-z_2 \left(a_2 \left(z_3
   \left(b_3-c\right)+b_2 \left(z_3-1\right)\right)+c (c+1)\right)}   {c (c+1)\left(z_2-1\right)}
   \nonumber\\ &&
   +\frac{a_2 b_3 z_3-a_2 c z_3+c^2+c}
   {c (c+1)\left(z_2-1\right)}
+\frac{z_1 \left(a_2 z_3+a_1-c\right)-a_2 z_3+c}{c (c+1)} \theta_1
 \nonumber \\ &&
-\frac{z_2 \left(a_2-b_2 z_3-b_3
   z_3+b_2+c z_3-2 c-1\right)-a_2 z_3-b_1 z_1 \left(z_2-1\right) \left(z_3-1\right)}{c (c+1)\left(z_2-1\right)}\theta_2
    \nonumber \\ &&
   +\frac{b_3 z_3+c+z_3}{c (c+1)\left(z_2-1\right)}\theta_2
 \nonumber \\ &&
  + \frac{z_2 \left(z_3 \left(b_3-c\right)+b_2 \left(z_3-1\right)+c\right)+b_1 z_1
      \left(z_2-1\right) \left(z_3-1\right)-b_3 z_3+c z_3-c}{c (c+1) \left(z_2-1\right)}\theta_3
 \nonumber \\ &&
\frac{\left(z_1
  + \left(z_2-1\right)-z_2\right) \left(z_3-1\right)}{c (c+1) \left(z_2-1\right)}\theta_1\theta_2
  -\frac{-z_3z_1+z_1+z_3}{c^2+c}\theta_1\theta_3
\Biggr]
\nonumber \\ &&
\times
F_S(a_1+1,a_2;b_1,b_2,b_3+1;c+2;z_1,z_2,z_3).
\end{eqnarray}

Also the hypergeometric function $F_S$ is not built into the current version
of {\tt Mathematica}, whereas our package implements the series representation 
of the hypergeometric function $F_S$, Eq.~(\ref{FS:series}).
Again, this series representation is suitable for numerical checks of the
results of the differential reduction and the corresponding examples are
gathered in the file {\tt example-FsFunction.m} available from~\cite{bytev:hyper}.

%
%
%

\section{On the construction of coefficients of the $\ep$-expansion of Horn-type
         hypergeometric functions}

For physical applications, the construction of analytical coefficients
of the Laurent expansion of hypergeometric functions around particular values 
of parameters (integer, half-integer, rational) is necessary \cite{nested, nested2, half-integer}. 
Many efforts have been made in the past in attempts to write results of the Laurent expansion 
in terms of multiple polylogarithms \cite{Goncharov}.
There are a number of different though entirely equivalent ways to describe the hypergeometric functions \cite{Gelfand}:
\begin{itemize}
\item[(i)] as a multiple series;
\item[(ii)] as a solution of a system of differential equations;
\item[(iii)] as an integral of the Euler type.
\end{itemize}
Each of these approaches can be used for
the construction of the Laurent expansion of hypergeometric functions
and each of them has some technical advantage or disadvantage
in comparison with the other ones.

The most universal technique which does not depend on the order of differential equation 
is based on the algebra of multiple sums \cite{nested,nested2}.
For the hypergeometric functions, where algorithms \cite{nested,nested2} are applicable, 
the result of the $\ep$-expansion are automatically written in terms of multiple polylogarithms.
The algorithms \cite{nested} have been implemented in a few packages \cite{nested3}.
However there are some technical problems with the extension of this
approach to rational values of parameters and its application specific classes of hypergeometric functions.

The differential equation approach \cite{expansion:1}
allows to analyze arbitrary sets of parameters simultaneously
and to construct the solution in terms of iterated integrals, but for any hypergeometric function
the Pfaff system of differential equations should be constructed.

Purely numerical approaches \cite{NumExp}
can be applied to for arbitrary values of the parameters. 
However this technique typically does not produce stable numerical result 
around region of singularities of hypergeometric function. 

The integral representation is well suited for the construction 
of the $\ep$-expansion for a limited number of types of hypergeometric functions \cite{integral}.

For hypergeometric functions considered in this paper, 
the physically interesting cases correspond to 
the construction of the $\ep$-expansion for the hypergeometric function 
$F_D$ around integer values of parameters \cite{FD}, 
and for $F_S$ around half-integer values of parameters \cite{fjt}, respectively.
For the  Appell functions $F_D$, the following theorem about 
structure of coefficients of all-order  $\ep$-expansion is valid: 

{\bf Theorem 2:} \\
\ {\it
The $\ep$-expansions of the Appell hypergeometric function $F_D$ of r-variables
and its first  derivative have the following structure:
\begin{subequations}
\label{FD-Theorem}
\begin{eqnarray}
&& 
F_D(a\ep; \{b_i \ep \}; 1+c \ep;\vec{z}) \;,
= 1 + a \sum_{k=2}^\infty r_k \Phi_k\left(\{1,\vec{z},\frac{z_i}{z_j}\}\right) \ep^k \;, 
\\ && 
z_i \frac{d}{dz_i}F_D(a\ep; \{b_j \ep \}; 1+c \ep;\vec{z})  
= a \sum_{k=2}^\infty d_k \rho_{k-1}\left(\{1,\vec{z},\frac{z_i}{z_j}\}\right) \ep^k \;, 
\end{eqnarray}
\end{subequations}
where $a,\{b_j\},c$ are an arbitrary numbers,
$\ep$ is an arbitrary small parameter,
$r_k$ and $d_k$ are the polynomial in $a,\{b_j\},c$
and $\Phi_{k}$ and $\rho_{k}$
include only  multiple polylogarithms of weight ${\bf k}$.}
\\

The proof of this theorem is based on the explicit construction 
of the iterative solution of the system of differential equations.

Up to functions of weight ${\bf 3}$ the explicit value of the coefficients are the following
\begin{eqnarray}
F^{(r)}_D(a\ep; \{b_i \ep \}; 1+c \ep; \vec{z}) 
& = & 
1 + \sum_{k=2}^\infty \omega_0^{(k)}(\vec{z}) \ep^k  \;, 
\nonumber \\ 
z_i \frac{d}{dz_i}F^{(r)}_D(a\ep; \{b_k \ep\}; 1+c \ep;\vec{z})  
& \equiv  & 
z_i \frac{a b_i \ep^2}{1+c\ep}F^{(r)}_D(1+a\ep; 1+b_i\ep, \{ b_j \ep\}; 2+c \ep;\vec{z})  
\nonumber \\  
& = & 
\sum_{k=2}^\infty \omega_i^{(k)}(\vec{z}) \ep^k  \;, 
\end{eqnarray}
where 
\begin{eqnarray}
\frac{\omega_0^{(2)}(\vec{z})}{a} &  = & \sum_{j=1}^r b_j \Li{2}{z_j}
\\
\frac{\omega_0^{(3)}(\vec{z})}{a} &  = &
\sum_{j=1}^r (a-c+b_j) b_j \Snp{1,2}{z_j}
- c \sum_{j=1}^r b_j \Li{3}{z_j}
\nonumber \\ && 
\!+\! \sum_{i<j=1; i \neq j}^r
b_i b_j
\left[
\Li{1,2}{\frac{z_i}{z_j},z_j}
\!+\!
\Li{1,2}{\frac{z_j}{z_i},z_i}
\right] \;,  
\end{eqnarray}
and 
\begin{eqnarray}
\frac{\omega_j^{(2)}(\vec{z})}{a} &  = & -b_j \ln(1-z_j) \;,  
\\
\frac{\omega_j^{(3)}(\vec{z})}{a} &  = &
b_j \Biggl[ 
\frac{1}{2} (a+b_j-c) \ln^2 (1-z_j)
- c  \Li{2}{z_j}
\!+\! \sum_{k=1; k \neq j}^r b_k \Li{1,1}{\frac{z_k}{z_j},z_j}
\Biggr]
\;,
\end{eqnarray}
where 
$
j=1,.\cdots, r 
$
and 
\begin{eqnarray}
\Li{k_1,k_2, \ldots, k_n}{x_1, x_2, \ldots, x_n} 
& = & \sum_{0< m_1 <  m_2 < \cdots < m_{n-1} < m_n}^\infty \frac{x_1^{m_1}}{m_1^{k_1}} \frac{x_2^{m_2}}{m_2^{k_2}}
\times\cdots\times \frac{x_n^{m_n}}{m_n^{k_n}}\;,
\label{GP}
\end{eqnarray}
denotes a multiple polylogarithm \cite{Goncharov}.

Starting from $\omega_j^{(4)}(\vec{z})$,  
the result of the $\ep$-expansion has more a complicated form.
Since the $F_D$ hypergeometric function is symmetric with respect to 
the exchange of the variables $(b_i,z_i) \leftrightarrow (b_j,z_j)$
we present here result only  for $\omega_1^{(4)}(\vec{z})$.
For case of two variables, we have: 
\begin{eqnarray}
&& 
\frac{\omega_1^{(4)}(x,y)}{ab_1}  =  
  \Delta_{b_1}^2 \Li{1,1,1}{1,1,x}
+ \left[ a b_1  - c \Delta_{b_1} \right] \Li{2,1}{1,x} 
- c \Delta_{b_1} \Li{1,2}{1,x} 
+ c^2 \Li{3}{x} 
\nonumber \\ && 
+ b_2 \Delta_{b_1} \Li{1,1,1}{\frac{y}{x},1,x}
+ b_2 \Delta_{b_2} \Li{1,1,1}{1,\frac{y}{x},x}
+ b_1 b_2 \Li{1,1,1}{\frac{x}{y},\frac{y}{x},x}
\nonumber \\ && 
- c b_2 \Li{1,2}{\frac{y}{x},x} 
+ (a-c) b_2 \Li{2,1}{\frac{y}{x},x} 
\;, 
\label{F1:1:4}
%
%
%
\end{eqnarray}
where 
$$
\Delta_{b} = a \!-\! c \!+\! b \;.
$$
Eq.~(\ref{F1:1:4}) coincides with results of \cite{nested3}.

For $r$-variable case $(r \geq 3)$ the iterative solution is:
\begin{eqnarray}
&&
\frac{1}{a b_1 b_r}
\left[
\omega_1^{(4)}(z_1,\cdots,z_{r-1},z_r)
-
\omega_1^{(4)}(z_1,\cdots,z_{r-1},z_r=0)
\right]
\nonumber \\ &&
=
(a-c)
\left[
  \Li{1}{\frac{z_r}{z_1}}\Li{1,1}{1,z_1}
- \Li{1,1,1}{1,z_1,\frac{z_r}{z_1}}
- \Li{1,2}{1,z_r}
\right]
\nonumber \\ &&
+ b_r  \Li{1,1,1}{1,\frac{z_r}{z_1},z_1}
- c   \Li{1,2}{\frac{z_r}{z_1},z_1}
\nonumber \\ &&
+  b_1
\left[
\Li{1,1,1}{\frac{z_1}{z_r},\frac{z_r}{z_1},z_1}
+
\Li{1,1,1}{\frac{z_r}{z_1},1,z_1}
\right]
\nonumber \\ &&
+ \sum_{j=1; j \neq 1; j \neq r}^{r} b_j
\left[
\Li{1,1,1}{\frac{z_j}{z_r},\frac{z_r}{z_1},z_1}
+
\Li{1,1,1}{\frac{z_r}{z_j},\frac{z_j}{z_1},z_1}
\right]
\;,
\end{eqnarray}
and further simplifications can be done by help of the stuffle relation:
\begin{eqnarray}
  \Li{1}{\frac{z_r}{z_1}}\Li{1,1}{1,z_1}
& = &
  \Li{1,1,1}{1,z_1,\frac{z_r}{z_1}}
+ \Li{1,1,1}{1,\frac{z_r}{z_1},z_1}
+ \Li{1,1,1}{\frac{z_r}{z_1},1,z_1}
\nonumber \\ &&
+ \Li{1,2}{1,z_r}
+ \Li{2,1}{\frac{z_r}{z_1},z_1} \;.
\end{eqnarray}

As an example for the application of these coefficients,
let us evaluate the function $F_D^{(r)}(1;1,\cdots,1;2;z_1,\cdots,z_r)$. 
Using the differential reduction algorithm, we get the following expression: 
\begin{eqnarray}
&& 
F_D^{(r)}(1\!+\!a;\{1\!+\!b_k\};1\!+\!c;z_1,\cdots,z_r)
= 
\nonumber \\ && 
\frac{c}{a} 
\sum_{i=1}^r \frac{z_i^{r-2}}{\Pi_{j=1; j \neq i}^r{(z_i-z_j)}} \frac{\theta_i}{b_i}
F_D^{(r)}(a;\{b_k\};c;z_1,\cdots,z_r) \;.
\end{eqnarray}
After replacing, $a \to a\ep, b_k \to b_k \ep, c \to 1+c \ep $
and taking limit $\ep \to 0$, we get (see \cite{FD:integer}):
\begin{eqnarray}
F_D^{(r)}(1;\{1\};2;z_1,\cdots,z_r)
& = &  
\sum_{i=1}^r \frac{z_i^{r-2}}{\Pi_{j=1; j \neq i}^r{(z_i-z_j)}} \frac{\omega_i^{(2)}(\vec{z})}{a b_i}  
\nonumber \\
& \equiv & 
- \sum_{i=1}^r \frac{z_i^{r-2}}{\Pi_{j=1; j \neq i}^r{(z_i-z_j)}} \ln(1-z_i) 
\;.
\end{eqnarray} 

In the end, 
we also present the theorem~\footnote{The proof is straightforward
and similar to the technique described in \cite{expansion:1,theorem}.} 
about structure of the $\ep$-expansion for two other Appell hypergeometric functions
in two variables:\\
\indent
{\bf Theorem 3:} \\
\ {\it
The $\ep$-expansions of the Appell hypergeometric functions $F_2$ and $F_3$ 
of 2-variables and their derivatives have the following structure:
\begin{subequations}
\label{F2-Theorem}
\begin{eqnarray}
&& 
F_{2}(a\ep, \{b_i \ep \}; \{1+c_i \ep\};z_1,z_2) 
= 1 + \sum_{k=2}^\infty r_k \Phi_k\left(\{1,\vec{z},\frac{z_i}{z_j}\}\right) \ep^k \;, 
\\ && 
z_i \frac{d}{dz_i} F_{2}(a\ep, \{b_i \ep \}; \{1+c_i \ep\};z_1,z_2) 
= \sum_{k=2}^\infty d_k \rho_{k-1}\left(\{1,\vec{z},\frac{z_i}{z_j}\}\right) \ep^k \;, 
\\ && 
z_1 z_2  \frac{d}{dz_1} \frac{d}{dz_2} 
F_{2}(a\ep, \{b_i \ep \}; \{1+c_i \ep\};z_1,z_2)  
= \sum_{k=3}^\infty q_k \sigma_{k-2}\left(\{1,\vec{z},\frac{z_i}{z_j}\}\right) \ep^k \;, 
\end{eqnarray}
\end{subequations}
\begin{subequations}
\label{F3-Theorem}
\begin{eqnarray}
&& 
F_{3}(a_1\ep, a_2 \ep; b_1 \ep, b_2 \ep; 1+c \ep;z_1,z_2) 
= 1 + \sum_{k=2}^\infty \tilde{r}_k \tilde{\Phi}_k\left(\{1,\vec{z},\frac{z_i}{z_j}\}\right) \ep^k \;, 
\\ && 
z_i \frac{d}{dz_i} F_{3}(a_1\ep, a_2 \ep; b_1 \ep, b_2 \ep; 1+c \ep;z_1,z_2) 
= \sum_{k=2}^\infty \tilde{d}_k \tilde{\rho}_{k-1}\left(\{1,\vec{z},\frac{z_i}{z_j}\}\right) \ep^k \;, 
\\ && 
z_1 z_2  \frac{d}{dz_1} \frac{d}{dz_2} 
F_{3}(a_1\ep, a_2 \ep; b_1 \ep, b_2 \ep; 1+c \ep;z_1,z_2) 
= \sum_{k=4}^\infty \tilde{q}_k \tilde{\sigma}_{k-2}\left(\{1,\vec{z},\frac{z_i}{z_j}\}\right) \ep^k \;,
\end{eqnarray}
\end{subequations}
where $\{a_i, b_j\},c$ are arbitrary numbers,
$\ep$ is an arbitrary small parameter,
$r_k, d_k, q_k$ and 
$\tilde{r}_k, \tilde{d}_k, \tilde{q}_k$  
are polynomials in $\{a_i, b_j\},c$, 
$\Phi_{k}$, $\rho_{k}, \sigma_k$
and 
$\tilde{\Phi}_{k}$, $\tilde{\rho}_{k}, \tilde{\sigma}_k$
include only multiple polylogarithms of weight ${\bf k}$.}

{\bf Example}: 
\begin{eqnarray}
&& 
F_{3}(a_1\ep, a_2 \ep; b_1 \ep, b_2 \ep; 1+c\ep; z_1,z_2) 
 =   
1 
+ 
\ep^2 \sum_{j=1}^2 a_j b_j \Li{2}{z_j} 
\nonumber \\ &&
+ 
\ep^3 \sum_{j=1}^2 a_j b_j 
\left\{ 
(a_j+b_j-c) \Snp{1,2}{z_j}
- c \Li{3}{z_j} 
\right\} 
+ O (\ep^4) \;.   
\end{eqnarray}

The $\ep$-expansion for $F_S$ around half-integer values of parameters will be considered 
in another paper.

\section{Conclusion }
\label{conclusion}
The differential-reduction algorithm for Horn-type hypergeometric functions 
allows one to compare different functions in this class whose values 
for the parameters differ by integers.
This proceeds in an entirely algorithmic manner suitable for automation 
in a computer algebra system.
In this paper, we have presented the {\tt Mathematica}-based 
programs {\bf FdFunction}  and {\bf FsFunction} 
for the differential reduction of the generalized hypergeometric function $F_D$ of $r$ variables
and the Lauricella-Saran hypergeometric function $F_S$ of three variables. 
Both functions are related with one-loop massive Feynman diagrams 
and both belong to the class of Horn-type hypergeometric function of order two.

For hypergeometric function $F_S$, Eq.~(\ref{FS:series}),  
we have presented a detailed analysis 
of the dimension of the solution space (see {\bf Theorem 1 } in Section~\ref{FS:section}).  
Regarding the expansion of the hypergeometric functions in a small coefficient $\ep$ 
around specific values of their parameters, 
we have proven two theorems $({\bf Theorem 2}$ and ${\bf Theorem 3})$.
These concern the structure of the all-order $\ep$-expansion of 
the hypergeometric functions, $F^{(r)}_D,F_2,F_3$ around integer values of parameters.
As illustration, the first three coefficients (up to functions of weight {\bf3})
have been constructed explicitly.
\vspace{5mm}
\noindent
\subsubsection*{Acknowledgments} 
We are grateful to T.~Riemann, O.~Tarasov, and O.~Veretin for useful discussions.
This work was supported in part by the 
Deutsche Forschungsgemeinschaft in SFB/TR 9, and by the Heisenberg-Landau Program. 
V.V.B. was supported in part by the 
Russian Foundation for Basic Research RFFI through Grant No.~12-02-31703.


\begin{center}
{\Large\bf Appendix}
\end{center}

\appendix

For completeness, we present also the integral representation of functions $F_D$ and $F_S$:
\begin{eqnarray}
&&\hspace*{-8mm}
\frac{\Gamma(a) \Gamma(c-a)}{\Gamma(a)}
F_D(a;b_1, \cdots, b_k;c_1, \cdots, c_k;z_1, \cdots, z_k) = 
\nonumber \\ &&
=
\int_0^1
du
u^{a-1} (1-u)^{c-a-1}
(1-uz_1)^{-b_1}
\cdots
(1-uz _n)^{-b_n} \;,
\label{FD:integral}
\\
&&\hspace*{-8mm}
\frac{\Gamma(a_1) \Gamma(a_2) \Gamma(c-a_1-a_2)}{\Gamma(c)}
F_S(a_1;a_2;b_1,b_2,b_3;c;z_1,z_2,z_3) = 
\label{FS:integral}
\\ &&
=
\int_{u,v \geq 0}
\int_{u+v \leq 1}
du dv 
u^{a_1-1} 
v^{a_2-1} 
(1-u-v)^{c-a_1-a_2-1}
(1\!-\!uz_1)^{-b_1}
(1\!-\!vz_2)^{-b_2}
(1\!-\!vz_3)^{-b_3}
\nonumber 
\;.
\end{eqnarray}


\end{document}